\documentclass[10pt, two column, twoside]{IEEEtran}
\usepackage{color}
\usepackage[colorlinks, linkcolor=color1, anchorcolor=blue, citecolor=color1]{hyperref}

\usepackage{amssymb}
\usepackage{amsmath}
\usepackage[lined,boxed,commentsnumbered, ruled]{algorithm2e}
\usepackage{mathrsfs}
\usepackage{algorithmic}
\usepackage{bm}
\usepackage{hhline}

\usepackage{pgfplots}
\usepackage{tikz}
\usetikzlibrary{arrows}
\usepackage{graphicx,booktabs,multirow}
\usepackage{subcaption}
\usepackage{graphicx}
\usepackage{makecell}
\usepackage{verbatim}

\definecolor{colorhkust}{RGB}{20,43,140}
\definecolor{colortsinghua}{RGB}{116,52,129}
\definecolor{color1}{RGB}{128,0,0}

\date{}

\begin{document}

\title{A Survey on Integrated Sensing, Communication, and Computation}
\author{Dingzhu Wen, \IEEEmembership{Member, IEEE}, Yong Zhou, \IEEEmembership{Senior Member, IEEE}, Xiaoyang Li, \IEEEmembership{Member, IEEE}, Yuanming Shi, \IEEEmembership{Senior Member, IEEE}, Kaibin Huang, \IEEEmembership{Fellow, IEEE}, and Khaled B. Letaief, \IEEEmembership{Fellow, IEEE}
        \thanks{D. Wen, Y. Zhou, and Y. Shi are with School of Information Science and Technology, ShanghaiTech University, Shanghai 201210, China. (e-mail: \{wendzh, zhouyong, shiym\}@shanghaitech.edu.cn).}
	   \thanks{X. Li is with Shenzhen Research Institute of Big Data, The Chinese University of Hong Kong-Shenzhen, Guangdong, 518172, China (e-mail:lixiaoyang@sribd.cn).}
	   \thanks{K. Huang is with Departement of Electrical and Electronic Engineering, The  University of  Hong Kong, Hong Kong (e-mail: huangkb@eee.hku.hk).}
        \thanks{K. B. Letaief is with Department of Electronic and Computer Engineering, The Hong Kong University of Science and Technology, Hong Kong (e-mail: eekhaled@ust.hk).}
}

\maketitle

\begin{abstract}
The forthcoming generation of wireless technology, 6G, promises a revolutionary leap beyond traditional data-centric services. It aims to usher in an era of ubiquitous intelligent services, where everything is interconnected and intelligent. This vision requires the seamless integration of three fundamental modules: Sensing for information acquisition, communication for information sharing, and computation for information processing and decision-making. These modules are intricately linked, especially in complex tasks such as edge learning and inference. However, the performance of these modules is interdependent, creating a resource competition for time, energy, and bandwidth. Existing techniques like integrated communication and computation (ICC), integrated sensing and computation (ISC), and integrated sensing and communication (ISAC) have made partial strides in addressing this challenge, but they fall short of meeting the extreme performance requirements.  To overcome these limitations, it is essential to develop new techniques that comprehensively integrate sensing, communication, and computation. This integrated approach, known as Integrated Sensing, Communication, and Computation (ISCC), offers a systematic perspective for enhancing task performance.  This paper begins with a comprehensive survey of historic and related techniques such as ICC, ISC, and ISAC, highlighting their strengths and limitations. It then discusses the benefits, functions, and challenges of ISCC. Subsequently, the state-of-the-art signal designs for ISCC, along with network resource management strategies specifically tailored for ISCC are explored. Furthermore, this paper discusses the exciting research opportunities that lie ahead for implementing ISCC in future advanced networks, and the unresolved issues requiring further investigation. ISCC is expected to unlock the full potential of intelligent connectivity, paving the way for groundbreaking applications and services.
\end{abstract}

\section{Introduction}
\begin{figure*}[]
\centering
\includegraphics[width=1\textwidth]{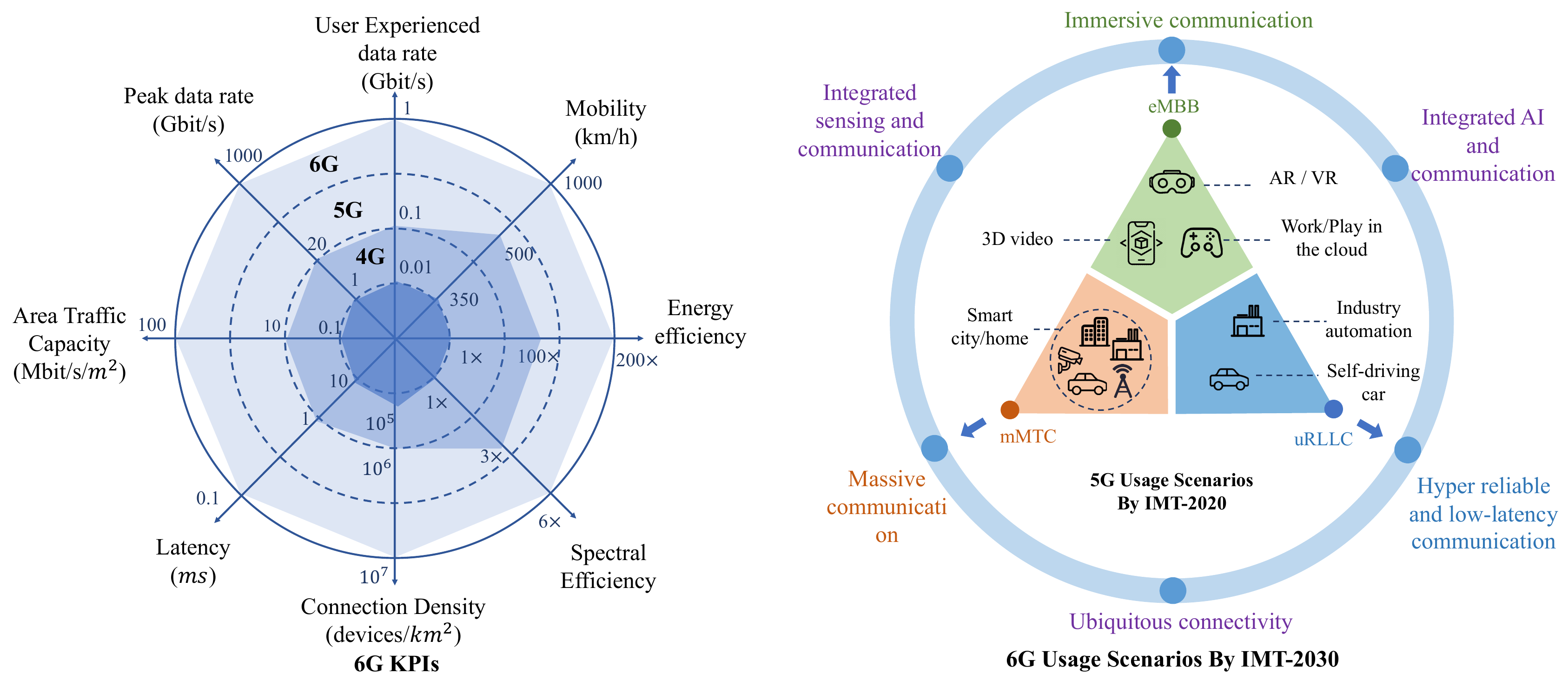}
\caption{Evolution of KPIs and usage scenarios of 6G.}
\label{Fig:KPIs_and_Scenarios}
\end{figure*}

Across the boundaries of fifth-generation wireless technology (5G), the next generation of wireless communications, 6G, is expected to go far beyond conventional communications of connected people and things to connected intelligence or intelligence of everything \cite{letaief2019roadmap,saad2020vision,huawei,10637271}. As announced by IMT-2030 three additional usage scenarios of integrated artificial intelligence (AI) and communications, integrated sensing and communication (ISAC), and ubiquitous connectivity, will be fostered in 6G, as well as three enhanced usage scenarios, i.e., immersive communication, massive communication, and hyper reliable and low-latency communication, which evolve from the scenarios of enhanced mobile broadband (eMBB), massive machine-type communication (mMTC), and ultra-reliable low-latency communication (URLLC) in 5G \cite{IMT2030}. To facilitate the implementation of these scenarios, higher key performance indicators (KPIs) are designed in 6G \cite{huawei,jiang2021the}. For example, the user-experienced data rate should achieve 1 Gbit/s and the connection density should be $10^7$ devices/km$^2$. Fig. \ref{Fig:KPIs_and_Scenarios} depicts the evolution of KPIs and usage scenarios. To achieve these KPIs, novel theories and techniques such as electromagnetic information theory \cite{zhu2024electromagnetic}, semantic communications \cite{yang2023semantic}, edge AI \cite{letaief2022edge}, near-field communications \cite{10558818}, and new hardware like frequency-selective surfaces \cite{10242165}, are proposed. Moreover, the realization of connected intelligence demands the implementation of massive intelligent applications at the network edge such as autonomous vehicles, smart cities, remote health care, and industrial automation \cite{letaief2022edge}. These applications call for the fusion of physical, biological, and cyber worlds, which naturally involves three modules, including sensing the environment for obtaining information, communication between different entities for information sharing, and computation on devices and servers for processing the information and making intelligent decisions \cite{wen2024integrated}. This gives rise to a brand-new research area called integrated sensing, communication, and computation (ISCC). In the sequel, the related partial integration technologies are first introduced, followed by the discussion of
the motivations of ISCC, the comparison of this survey with related works, and the summary of our contributions.
{

\subsection{Technologies of Partial Integration}

\subsubsection{Integrated Communication and Computation} 
Integrated communication and computation (ICC) refers to the joint design of communication and computation processes in tasks by allowing resource allocation between them to improve the system performance under a unified goal, such as energy/latency minimization and computation accuracy maximization. Since a dozen years ago, researchers in the wireless communication field have made efforts to develop ICC techniques at the network edge. Initially, mobile edge computing (MEC) is proposed, which aims at offloading partial or full on-device tasks to edge servers for enhancing the system performance e.g., energy efficiency and latency, via balancing the on-device communication and computation loads \cite{mao2017survey}. Building on MEC, edge AI has attracted considerable attention, which aims at conducting distributed model training and cooperative inference at the network edge to exploit the benefits of fast mobile data access, data privacy preservation, and so on \cite{letaief2022edge}. With the rapid advancement of microchip technology (the number of computer chips doubles every 18 to 24 months according to Moore's Law), massive computing powers are deployed on cloud, edge, and devices in wireless networks. Accordingly, a new ICC paradigm called computing power networks is proposed for coordinating and scheduling the ubiquitous computing resources to accomplish computationally intensive tasks, such as distributed training of large models \cite{tang2021computing,yukun2024computing}. Particularly, since the deployment of massive low-earth orbit (LEO) satellites is regarded as the solution for providing infrastructure-free areas with communication services \cite{liu2018space}, space computing power networks will be a promising research topic. Apart from the ICC schemes in the application layer, a physical layer technique called over-the-air computation (AirComp) has emerged in recent years \cite{zhu2019mimo}. It allows all transmitters to send their signals over the same wireless resource block and directly calculates a functional value of all signals at the receiver using the waveform superposition property, instead of decoding each transmitter's signal. As a result, the communication efficiency is significantly enhanced. Due to its merits, AirComp has been adopted as an emerging air-interface to support many applications, including wireless data aggregation \cite{wen2019reduced,li2019wirelessly,cao2020optimized,li2018air}, federated edge learning \cite{zhu2020broadband,yang2020federated,shao2021federated, 9502547}, and edge AI inference \cite{yilmaz2022over, wen2022taskaircomp, yang2023communication}.

\subsubsection{Integrated Sensing and Computation}

Integrated sensing and computation (ISC) refers to the joint design of sensing transceivers and the subsequent sensing echo signal processing algorithms. The development of ISC techniques is significant since the realization of connected intelligence calls for deploying massive sensors in the network to sense the environment \cite{feng2021joint}. In this paper, we categorize ISC technologies into three types. The first is wireless (radar) sensing, which was originally developed in the military field for detecting aircraft before World War II. By carefully designing a transmit waveform, e.g., frequency modulation continuous wave (FMCW), and processing the corresponding echo signals with the algorithms of sampling, fast Fourier transform (FFT), correlation calculation, filtering, etc., the distance and velocity of the target can be detected \cite{liu2020joint}. Other waveforms such as orthogonal frequency division multiplexing (OFDM) waves, novel techniques like multiple antennas, and new frequency bands such as millimeter wave bands, can be used for enhancing the detection performance \cite{liu2020joint}. In addition, the utilization of wireless sensory data to train AI models for supporting various intelligent tasks such as human motion recognition and natural scene analysis, has become a research trend \cite{liu2020joint}. Radar sensing is widely used in many military applications including military target detection, tracking, and recognition, and civil applications such as disaster assessment, land use and classification, and weather detection \cite{liu2020joint}. However, a single modality of wireless sensing is not able to obtain sufficient information for complicated intelligent tasks \cite{NEURIPS2021_5aa3405a}. To this end, multi-modal sensing (including radar, camera, microphone, etc.) technology is developed. Benefiting from the rapid development of AI, machine learning algorithms are generally used for multi-modal sensory data processing, including representation, translation, alignment, and fusion  \cite{8269806,song2024multi}. The typical use cases of multi-modal sensing include media description applications like image and video captioning,  speech recognition and synthesis, human action classification, multimedia event detection, and emotion recognition \cite{song2024multi}. Moreover, mobile devices equipped with several types of sensors such as accelerometers, GPS, cameras, and microphones, have experienced exponential growth in the last several decades. This leads to a new paradigm called mobile crowdsensing, which collects massive and multi-domain sensory data from mobile devices for providing intelligent and people-centric services \cite{capponi2019survey}. The applications of mobile crowdsensing include accidents and natural disasters prevention, waste-recycling operations,  air quality monitoring, and so on \cite{capponi2019survey}.

\subsubsection{Integrated Sensing and Communication}

ISAC jointly designs the sensing and communication modules in the system, which saves hardware space and improves 
resource utilization by sharing hardware (e.g., transceivers) and network resources (e.g., time, spectrum, and energy), as well as making sensing and communication serve as each other's supporting techniques to enhance their respective performance, so-called sensing for communication and communication for sensing.  Recently, ISAC has attracted extensive research focuses and is regarded as one usage scenario of 6G by IMT-2030 \cite{IMT2030}. There are three ISAC paradigms, corresponding to the three ISC technologies. The first is integrated wireless sensing and communication \cite{liu2022integrated, cui2021integrating}. By equipping dual-functional radar communication (DFRC) systems on devices, integrated wireless sensing and communication enjoy the advantages of saving hardware space and sharing spectrum resources. One main challenge lies in dealing with co-channel interference. Particularly, Wi-Fi sensing that utilizes Wi-Fi signals for detection, recognition, and estimation \cite{hernandez2022wifi}, has been a hot topic in this area, as the Wi-Fi equipment is widely deployed and no extra hardware is needed \cite{chen2024deep}. However, complete and fine-grained characteristics of both sensing target and wireless channels cannot be obtained from only the uni-modal electromagnetic wave.  To this end, the second paradigm, i.e., integrated multi-modal sensing and communication, is proposed \cite{cheng2024intelligent}. One of its main functions is sensing for communication, where multi-modal sensory data is used to assist channel state information (CSI) acquisition. The other is communication for sensing, where communication plays a key role in multi-agent multi-modal sensing systems by enabling data fusion and coordination. For the third paradigm, i.e., integrated mobile crowdsensing and communication, current research focuses on utilizing existing communication techniques for collecting massively distributed sensory data. Thus, it remains a potential area for future investigation. Typical use cases of ISAC include simultaneous localization and mapping and vehicle platooning in vehicular networks, detection of human presence, proximity, and activities in smart homes, UAV monitoring and management, human-computer interaction applications, and so on \cite{cui2021integrating}.

\subsection{Motivations of ISCC}\label{Sect:ISCC_Motivation}

Future wireless networks are expected to complete complicated tasks to provide intelligent services such as smart cities, auto-driving, virtual/augmented reality, and smart manufacturing \cite{letaief2019roadmap,saad2020vision}. These tasks generally involve three modules, i.e., information perception (sensing), information transmission (communication), and information processing for intelligent decision-making (computation) \cite{wen2024integrated, shi2023task, zhu2023pushing}. Existing techniques, i.e., ISC, ICC, and ISAC as mentioned before, make efforts to support a part of intelligent tasks with loose performance requirements. However, they fail to achieve the extreme performance demands of real-time intelligent tasks, like ultra-high decision accuracy and ultra-low latency. Specifically, they have the following three shortages.
\begin{itemize}
    \item \emph{Low Resource Utilization}: In existing designs, network resources are shared and coordinated between only two of the sensing, communication, and computation modules, leading to low resource utilization. For example, in a sensing and AirComp supported edge-device cooperative inference task, joint resource allocation between sensing and AirComp can enhance resource utilization than the separate allocation scheme since the latter fails to adjust the resource budgets for sensing and AirComp according to the time-varying environment \cite{zhuang2024integrated}.

    \item \emph{Mismatch between Goals of the Overall Task and Each Module}: In existing designs, different modules are designed under separate design goals, e.g., sensing for obtaining high-quality sensory data, communication for throughput maximization, and computation for low energy cost and low latency. However, enhancing these goals is not necessarily beneficial or sometimes even harmful to the overall goal of the task \cite{lan2021semantic}.  For example, the transmission of unimportant data, like gradient vectors with zero values in federated learning tasks \cite{ren2020scheduling} and uninformative samples in active learning tasks \cite{wen2019overview,liu2020data}, consumes system resources (e.g., time, bandwidth, and energy) but makes little contribution to the task.
    \item \emph{Ignoring the Tight Coupling of Different Modules}: The sensing, communication, and computation modules are highly coupled in many tasks. For example, an online federated learning task involves sensing for data sample acquisition, on-device computation for local updating, and communication for local model uploading. The three modules compete for network resources (e.g., time and energy) but the convergence performance depends on the number of and the signal-to-noise ratio (SNR) of sensory data samples, the on-device computation speed, and the communication capacity in each round. Existing techniques ignore this tight coupling and design optimization algorithms from a partial perspective, leading to performance degradation.

\end{itemize}
By allowing resource sharing, coordination, and allocation among sensing, communication, and computation in a task-oriented way, i.e., under a unified task goal, ISCC can overcome      the above shortages. Moreover, ISCC can help achieve many 6G KPIs, including but not limited to enhancing spectrum efficiency, energy efficiency, and connection density, and reducing end-to-end task-completion latency by allowing resource sharing and coordination among sensing, communication, and computation and using intelligent algorithms like AI to process sensory data for enhancing communication performance. The details are described in Section \ref{Sect:6GFunctions}.


\subsection{Related Works and Our Contributions}

There have been several surveys and reviews related to the topic of ISCC \cite{zhu2023pushing,8944276,9694609,dai2022survey,zhao2023federated,wang2024integration}. The utilization of energy harvesting techniques to support sensing, communication, or computation tasks was surveyed in \cite{8944276}. The authors in \cite{9694609} surveyed the definition, preliminaries, and latest findings of integrated wireless sensing and communication and its application in IoT computation tasks. However, sensing, communication, and computation, or ISAC and computation, are considered as independent modules or tasks in the above two works, without investigating from the integration perspective. The other works focus on the adoption and application of ISCC to support certain techniques or scenarios. Specifically, the authors in \cite{dai2022survey} surveyed the separate maritime sensing, communication, and computing paradigms and the partial integration technologies for smart oceans such as ISAC and pointed out the challenges and applications of ISCC in maritime networks but put relatively little attention on surveying the ISCC techniques. The authors in \cite{zhao2023federated} provided a review of federated learning with an emphasis on the design from an ISCC perspective, without discussing the specific ISCC techniques. The authors in \cite{zhu2023pushing} further provided a survey about ISCC-supported edge AI, which investigated the resource allocation of three tasks, i.e., centralized edge learning, federated learning, and edge inference. The latest work surveyed the adoption and configuration of ISCC for supporting metaverse scenarios as well as the state-of-the-art techniques for ISCC implementation \cite{wang2024integration}.

Compared with existing surveys \cite{zhu2023pushing,8944276,9694609,zhao2023federated,wang2024integration}, we shall conduct a comprehensive survey about ISCC and provide new ideas, discussions, insights, and analysis. We start from introducing the three partially integrated technologies, i.e., ICC, ISC, and ISAC. The shortages of the three partially integrated technologies are particularly analyzed to motivate ISCC. Then, the benefits, functions, and challenges of ISCC are investigated to offer an overall understanding. Subsequently, two kinds of ISCC techniques, i.e., signal design and resource management, are surveyed and discussed, followed by providing the possible implementation of ISCC in future advanced networks and discussing the unresolved issues. The detailed contributions are listed as follows.
\begin{itemize}
\item \emph{Summary of Partially Integrated Technologies}: A comprehensive survey of the three partially integrated technologies, i.e., ISC, ICC, and ISAC, are provided, including their definitions, design goals, metrics, key techniques, and applications. In particular, the ISC and ISAC are categorized according to the three sensing types, i.e., wireless sensing, multi-modal sensing, and mobile crowdsensing. The connection and difference of the three categories are also provided. Two ICC technologies, i.e., MEC and AirComp, are introduced, which integrate computation and communication from the perspectives of task-level and signal-level respectively. These three partially integrated technologies provide the foundation for ISCC. 

\item \emph{Motivations, Benefits, Functions, and Challenges of ISCC}: The shortages of the three partially integrated technologies are first analyzed, i.e., low resource utilization, the mismatch between goals of the overall task and each module, and ignoring the tight coupling of different modules as mentioned. Then, the benefits of ISCC are discussed, including the sensing-communication-computation symbiosis to benefit each other of the three modules, as well as overcoming the aforementioned shortages of three partially integrated technologies for achieving higher resource utilization and task performance. Next, the functions of ISCC in 6G to support many KPIs and usage scenarios are provided, followed by a discussion of its challenges.


\item \emph{Signal Design for ISCC}: To enhance resource utilization, the signal-level designs in terms of the beamforming and waveform are first investigated. This is done to simultaneously improve the performance of sensing measured by e.g., the Cram{\'e}r-Rao bound (CRB), communication measured by SNR, and computation measured by the mean-squared-error (MSE). Three types of signal-level ISCC techniques are investigated. The first is called single-functional design, where independent signals are dedicated to sensing, communication, and computation functionalities. Its main challenge is the severe interference among the three kinds of signals. To alleviate the interference, the second type designs one signal for sensing and another for AirComp, called dual-functional signal design. The third is triple-functional signal design, which realizes the three functions in one signal, which trades increased design complexity for interference cancellation.

\item \emph{Network Resource Management for ISCC}:  There are two ISCC resource management paradigms. One is joint resource management for the coexistence of sensing, communication, and computation tasks. In this paradigm, a multi-objective optimization problem is formulated to guide the resource allocation among sensing, communication, and computation. The other is task-oriented ISCC resource allocation for complicated tasks where the three processes, i.e., sensing for information acquisition, computation for information processing, and communication for information transmission, are tightly coupled. A comprehensive survey for these two paradigms is conducted, followed by the discussion in terms of its benefits, challenges, and application scenarios. 

\item \emph{Looking to ISCC in the Future}: Given its huge potential, the ISCC-supported digital-twins networks, and the ISCC applications in two future advanced future networks, i.e., computing power networks, and space-air-ground integrated networks, are presented. Moreover, many critical issues of ISCC that have not been addressed yet are discussed in this work, including theoretical analysis tools, hardware design, protocols and standardization, security, privacy, robustness, and backward compatibility. 

\end{itemize}

\subsection{Organization}

\begin{figure*}[h]
\centering
\includegraphics[width=1\textwidth]{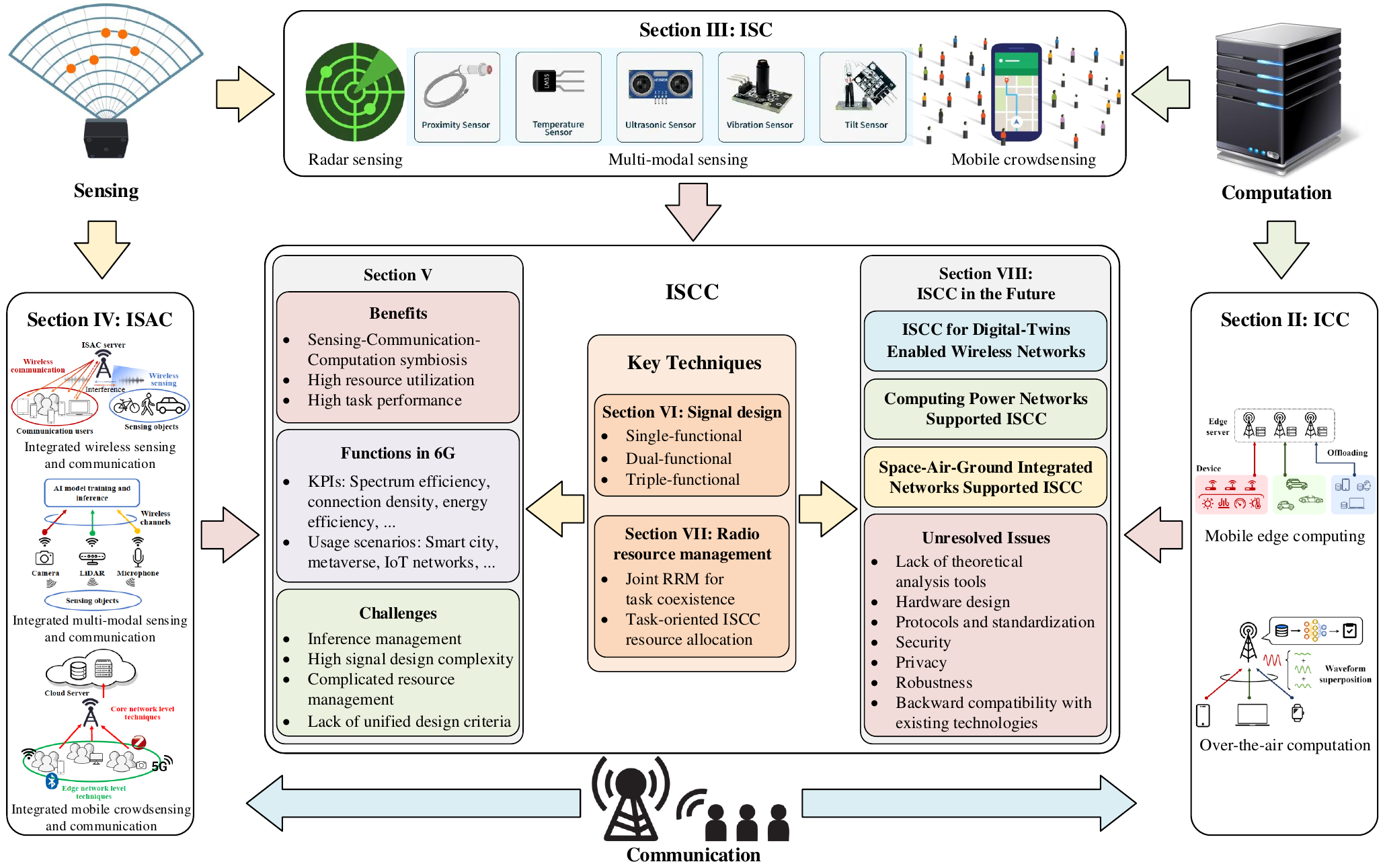}
\caption{The structure of this survey.}
\label{Fig:PaperStructure}
\end{figure*}

An overall structure of this survey together with the relations between different sections are presented in Fig. \ref{Fig:PaperStructure}. Specifically, Sections II, III, and IV conduct comprehensive surveys about the three existing partial technologies, which serve as the foundations and support of ISCC. Section V points out the shortages of the partial integration technologies and discusses the benefits, functions, and challenges of ISCC. Then, Sections VI and VII survey the state-of-the-art key techniques from the perspectives of signal design and network resource management, respetively, which manage to tackle the challenges of ISCC for enjoying its benefits. Subsequently, the future applications and unresolved issues are discussed in Section VIII. The detailed organization of this paper is listed below. 

\begin{itemize}
    \item Section II conducts a detailed survey of two ICC technologies, i.e., MEC and AirComp. 
    \item Three kinds of ISC technologies, i.e., wireless (radar) sensing, multi-modal sensing, and mobile crowdsensing, are discussed in Section III.
    \item In Section IV, a comprehensive investigation of ISAC is provided, including integrated wireless/multi-modal/crowd-sensing and communication techniques. 
    \item Section V discusses the benefits, functions, and challenges of ISCC.
	\item Section VI introduces three types of signal designs for ISCC, i.e., single-/dual-/triple-functional signal designs. Particularly, triple-functional signal design refers to using one signal to perform three functions of sensing, communication, and computation. 
    \item In Section VII, the network resource management paradigms for ISCC are discussed. One is joint RRM for task coexistence, which refers to resource allocation among different tasks of sensing, communication, and computation to achieve multiple goals simultaneously. The other is task-oriented ISCC resource allocation, which refers to allocating network resources among the sensing, communication, and computation modules of a task to achieve its customized goal. 
    \item In Section  VIII, the implementation of ISCC in future advanced networks is investigated in Section VIII, including ISCC for digital-twins enabled wireless networks, computing power networks supported ISCC, SAGINs supported ISCC. And the unresolved issues in ISCC that demand future research efforts are discussed.
    \item Section IX concludes the paper.
\end{itemize}


\section{Integrated Communication and Computation}

\subsection{Overview}

Emerging intelligent services over wireless networks often encompass both communication and computation processes. 
However, in conventional wireless networks, the communication and computation processes are separately designed, 
as the inherent connection between the transmitted data and subsequent service tasks is ignored. 
Such a separate design leads to inefficient communication and computation resource utilization and in turn degrades the execution performance of the underlying service tasks.
By treating the communication and computation processes as a unified module, ICC emerges as a promising technique to enhance the system performance, where the allocation of communication and computation resources are jointly optimized to achieve a common objective, e.g., energy/delay minimization and computation accuracy maximization, thereby making full use of limited resources to improve the system performance. 
In this section, we provide an overview of the existing studies on ICC, focusing on the efficient design of joint communication and computation, which is illustrated by presenting two promising examples, e.g., MEC and AirComp, as shown in Fig. \ref{Fig:ICC} and listed below. 

\begin{itemize}
	\item \textit{Mobile Edge Computing:}
	MEC intends to move the data transmission and processing operations to the network edge by exploiting the computing power and storage capabilities of edge servers and distributed devices, while the communication resources can be utilized to exchange for the computation resources via task offloading, thereby enabling ICC with joint communication and computation resource allocation. 

	\item \textit{Over-the-Air Computation:}
	AirComp aims at efficiently computing a target function by exploiting the co-channel interference, where the simultaneously transmitted signals can be naturally added over the air to realize a function computation during the communication process, thereby achieving ICC by exploiting the waveform superposition of concurrently transmitted signals.

\end{itemize}
\begin{figure*}[ht]
\centering
\includegraphics[width=1\textwidth]{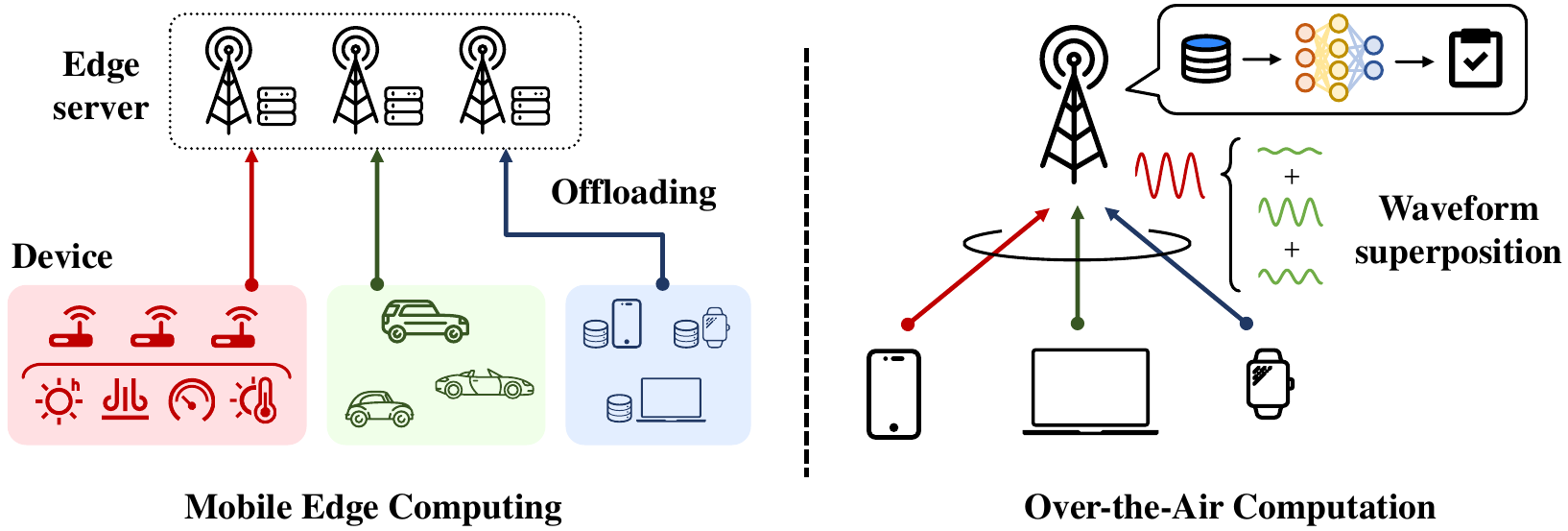}
\caption{Illustration of two typical examples of ICC, i.e., MEC and AirComp.}
\label{Fig:ICC}
\end{figure*}

\subsection{Mobile Edge Computing}
In the past few years, computation tasks over wireless networks are typically accomplished by resorting to cloud computing, which requires a large amount of data generated by devices to be collected and transmitted to a centralized server with high storage and computing capabilities, so as to perform data analysis and computation for meeting various requirements of different devices. 
However, with the rapid development of Internet of Things (IoT), the number of connected devices dramatically increases, leading to a severe communication burden on the wireless network and high computing pressure on the central server. 
As a result, cloud computing that typically relies on a remote central server is no longer suitable to support latency-critical services in future wireless networks with massive connectivity and limited radio resources. 
To address the above issues, MEC is emerging as a new computing paradigm thanks to the advancement of microchip computing power, which enables the edge server and mobile devices to execute computing tasks with their own computing and storage capabilities at the network edge \cite{mao2017survey}.

A typical MEC system consists of edge servers and mobile devices, where edge servers are generally co-located with access points (APs) and deployed in proximity to mobile devices.
Herein, the mobile devices are allowed to offload the computational tasks to edge servers for executing the tasks with high computing and low latency requirements.
To unleash the potential of MEC, numerous studies have explored task offloading design in MEC systems with single/multiple edge server(s) and multiple users. Specifically, the system design of MEC lies in the following aspects.

\begin{itemize}
	\item \textit{Joint Resource Allocation:}
	Even though the edge servers are capable of providing strong computing power to fulfill various service requirements for different devices, they still confront high computation and communication burdens when there are massive tasks from multiple devices that need to be accomplished due to the limited radio and computing resources of the MEC system.
	To overcome the above challenge, joint communication and computation resource allocation is required to enhance the system performance in terms of different objectives, e.g., execution delay and energy consumption. 
	For MEC in software-defined ultra-dense wireless networks with battery-limited devices, the task placement and resource allocation can be jointly optimized to minimize the average task duration \cite{chen2018task}.
	In remote areas lacking conventional edge/cloud infrastructures, a SAGIN-assisted MEC framework is developed in \cite{cheng2019space}, where task scheduling and resource allocation can be jointly optimized by developing a learning-based approach to reduce the execution delay.
	To mitigate the communication overhead, an efficient decentralized algorithm is developed in \cite{josilo2019selfish} to enable task offloading in a dense wireless network with selfish mobile devices. By using game-theoretical analysis, the computation costs, i.e., the weighted sum of delay and energy consumption, of different devices can be further balanced. 
	To efficiently support multi-user task offloading, non-orthogonal multiple access (NOMA) assisted MEC is proposed in \cite{9036885}, where the communication and computation resources are jointly optimized for minimizing the delay and energy consumption.


	\item \textit{Task Scheduling:}
	In real-world applications (e.g., autonomous driving, augmented reality, and video surveillance), tasks are not only delay-sensitive and computation-intensive but also dependent on each other, which demands sequential task processing.
	Hence, the output of one task becomes the input of the next one, resulting in a chain of dependencies among different tasks. Specifically, the computation tasks can be interdependent in various forms, i.e., sequential, parallel, and a combination of both. The interdependency makes it critical to schedule the offloading of such interdependent tasks in MEC systems, so as to ensure satisfactory application performance, which can be evaluated by the overall latency and energy consumption of all tasks in the application. 
	To this end, the interdependent tasks can be modeled as a directed acyclic graph, followed by developing a task offloading scheme for multiple interdependent computation tasks in MEC systems \cite{guo2024joint}. 
	By formulating a mixed-integer nonlinear programming problem, a joint scheduling and offloading design is proposed to effectively reduce the average completion time of interdependent tasks in MEC systems.
	Furthermore, the impact of interdependency among different tasks on the system reliability can be explored to facilitate the design of task scheduling as in \cite{liu2023dependent}, where a multi-priority task sequencing algorithm and a deep deterministic policy gradient-based learning algorithm are developed to minimize the system deadline violation ratio. 
	In industrial MEC networks, the assembly processing line involves multiple computing tasks with diverse priority and latency requirements. 
	This motivates the development of hybrid NOMA and orthogonal multiple access (OMA) based task offloading \cite{10609381}, where the selection of NOMA and OMA is based on the communication and computation latency requirements of the underlying tasks.


	\item \textit{Cooperative Computing:}
	The mobile devices in MEC systems generally offload their computation-intensive tasks to edge servers for low-latency task execution while reducing communication overhead.
	Nevertheless, edge servers may become unavailable due to factors such as limited coverage area and exhausted computing resources, resulting in new tasks offloaded by devices not being processed instantly or failing directly.
	To address the above challenge, multi-user cooperative computing emerges as a promising solution to enhance the reliability of MEC systems by alleviating edge server overload and exploiting computing resources distributed among mobile devices.
	For instance, a device-to-device (D2D)-assisted co-offloading framework for industrial IoT is proposed in \cite{dai2023task}, where an efficient learning-based algorithm is developed to minimize the weighted sum of task delay and migration cost without the need for complete offloading information. 
	In D2D-enabled MEC systems with time-varying idle resources and channel conditions, the task partitioning and parallel scheduling can be jointly optimized to achieve near-optimal delay performance with low computational complexity while ensuring fairness across various system states \cite{li2024joint}. 
	In addition, cooperative task offloading can also be applied in Internet of Vehicles (IoV) networks, where the peripheral vehicles with idle computing resources can be exploited to improve the task execution efficiency in dynamic vehicular environments \cite{chen2023multihop}. 	

\end{itemize}

\subsection{Over-the-Air Computation} \label{sec:aircomp}
In traditional communication systems, the signals concurrently transmitted by different devices over the same channel are deemed as co-channel interference, thus each of them demands individual and accurate reception at the receiver before the subsequent processing.
In the face of emerging intelligent services, achieving functional computation of gathered data becomes the key to implementing the corresponding applications, e.g., environmental monitoring and autonomous driving, while the accuracy and timeliness of such computation determine the quality of service.
In this sense, the conventional \textit{compute-after-communicate} strategy may become inefficient in a large-scale network with scarce resources due to the separation of communication and computation.
As a result, AirComp has attracted much attention to pursuing the fusion of communication and computation by harnessing the co-channel interference \cite{zhu2021overtheair, wang2022over, 10388035}.


Seeking to integrate communication and computation, AirComp regards the co-channel waveform superposition as a summation operation for synchronized concurrently transmitted signals, enabling \textit{compute-when-communicate} manner that allows the target function to be computed during the transmission process without need for one-by-one decoding at the receiver.
As all devices are required to occupy the same channel to enforce the computation for the target function, the number of radio resources utilized by AirComp is unrelated to the number of participating devices, thereby yielding a much higher spectral efficiency than conventional OMA approaches that allocate separated channels to each device.
Although power-domain NOMA can also enable concurrent transmission over the same channel by leveraging superposition coding and successive interference cancellation, the algorithmically complex and time-consuming interference cancellation operation is inevitable before performing the subsequent function computation.
As a result, compared with the NOMA counterpart, AirComp has the advantages of having low implementation complexity and saving the time for recovering each signal in terms of the functional computation at the cost of acceptable computation errors.


By implementing appropriate pre- and post-processing on transmit and receive signals, respectively, AirComp enables the receiver to compute a class of functions, namely nomographic function \cite{goldenbaum2015nomographic}, which is defined as $f(x_1, x_2, \dots, x_K) = \psi\left(\sum_{k = 1}^K \varphi_k(x_k)\right)$,
where $\psi(\cdot)$ denotes the post-processing function at the receiver and $\varphi_k(\cdot)$ denotes the pre-processing function for data $x_k \in \mathbb{R}$ at device $k$, $\forall \, k \in \{1, 2, \dots, K\}$.
Following the expression given in the above, AirComp can be decomposed into three stages \cite{zhu2019mimo}:

1) transmit data pre-processing at each device; 
2) summation of pre-processed data through the waveform superposition property of multiple-access channels;
   and 3) receive data post-processing at the access point (AP).
   Note that, although the signal superposition can only realize the operation of summation, AirComp is capable of computing non-linear functions \cite{abari2016over}, e.g., geometric mean and {Euclidean} norm, and even computing any continuous function by assembling multiple nomographic computation results \cite{goldenbaum2015nomographic}.



During the wireless propagation, the synchronized signals are performed with a weighted summation with respect to the channel coefficients, which may lead to a non-desired computation result.
Accordingly, the transceiver design is necessary for AirComp to induce an expected signal superposition over non-uniform fading channels.
Recently, many efforts have been made to induce signal alignment under different objectives for AirComp.
\begin{itemize}
	\item \textit{Data Fusion:}
	Data fusion is the fundamental objective of AirComp and much of the underlying research has been conducted from this aspect \cite{9163314}.
	For instance, to minimize the distortion of data fusion in single-input single-output (SISO) networks, the optimal transceiver design for AirComp under peak power constraints is developed in \cite{cao2020optimized}. 
	Meanwhile, the optimal design under sum-power constraint is obtained in \cite{zang2020overtheair} with closed-form expressions.
	By exploiting multi-antenna techniques, the beamforming design developed in \cite{zhu2019mimo, chen2018uniform, chen2018overtheair} validate the performance improvement that can be achieved via diversity gain and multiplexing gain to reduce fusion error \cite{chen2018uniform} and realize multi-modal fusion \cite{zhu2019mimo, chen2018overtheair}, respectively.
	To mitigate the reliance on perfect CSI, blind AirComp can be utilized to enable data fusion. One widely used method applies one-bit quantization at the transmitter, followed by a majority vote on the sign of the aggregated data at the receiver \cite{9641940}.
	

	\item \textit{Distributed Consensus:}
	Distributed consensus expects all devices in the network to reach a consensus on a certain message without the assistance of a central node, which can be accelerated by AirComp with its efficient data aggregation.
	Specifically, as demonstrated in \cite{nazer2011local}, multiple simultaneous average operations achieved by AirComp can accelerate the convergence rate of gossip-based consensus.
	Such an AirComp-assisted consensus framework is further extended to multi-cluster scenarios in \cite{6214113}, where the local averaging time is significantly reduced and the algorithmic convergence is established.
	Moreover, as demonstrated in \cite{lin2022distributed}, effective beamforming design is also essential to accelerate the distributed consensus of decentralized AirComp-based schemes. To facilitate accurate distributed consensus, achieving time and frequency synchronization at the receiver side is essential. To this end, timing advance \cite{mahmood2019time} and AirShare \cite{abari2015airshare} are effective techniques for guaranteeing time synchronization and eliminating carrier frequency offset, respectively.

	\item \textit{Model aggregation:}
	Federated learning (FL) has attracted much attention in machine learning areas due to its distributed computing structure and privacy-preservation property \cite{mcmahan2017communication}.
	Although transmitting model parameters reduce the communication overhead by a significant amount compared to transmitting raw data, scarce communication resources still limit the communication efficiency in wireless FL systems, which motivates the emergence of AirComp-based FL \cite{zhu2020broadband, yang2020federated, 9791337, sery2020analog}.
	As demonstrated in \cite{zhu2020broadband}, AirComp-based FL achieves much lower communication latency than the OFDMA-based scheme. 
	Device scheduling is an important issue for ensuring accurate model aggregation for FL \cite{yang2020federated, zhu2020broadband, 10355909}. 
	Moreover, compression and sparsification can also be integrated with AirComp to reduce the communication overhead of FL in resource-constrained wireless networks, as illustrated in \cite{mohammad2020machine}. AirComp can also be utilized to support multiple FL tasks in multi-cell wireless networks, where co-channel interference degrades the accuracy of model aggregation. This issue can be alleviated by developing a cooperative optimization framework that balances the learning performance across different FL tasks \cite{9791337}. 

\end{itemize}

\subsection{Discussion}
This section presents the principles and potentials of ICC, with a specific emphasis on elaborating two representative ICC techniques, i.e., MEC and AirComp. For both MEC and AirComp, the joint optimization of communication and computation resources is critical for enhancing the system performance. Moreover, task scheduling and cooperative computing are two key aspects of MEC, while the transceiver design for different scenarios is essential for AirComp. Various optimization-based algorithms have been developed to achieve efficient resource utilization in ICC. As 6G networks are expected to feature more complex architectures and larger network sizes, learning-based algorithms that have the potential to reduce computation complexity and enhance scalability for the joint communication and computation resource allocation deserve further investigation.

\section{Integrated Sensing and Computation}\label{Sect:ISC}

\subsection{Overview}
In this section, we provide a comprehensive survey about the technologies for both sensing signals acquisition and the corresponding processing methods for completing sensing tasks, giving its name ISC. As shown in Fig. \ref{Fig:SCMP}, there are three ISC technologies. 
\begin{figure*}[ht]
\centering
\includegraphics[width=1\textwidth]{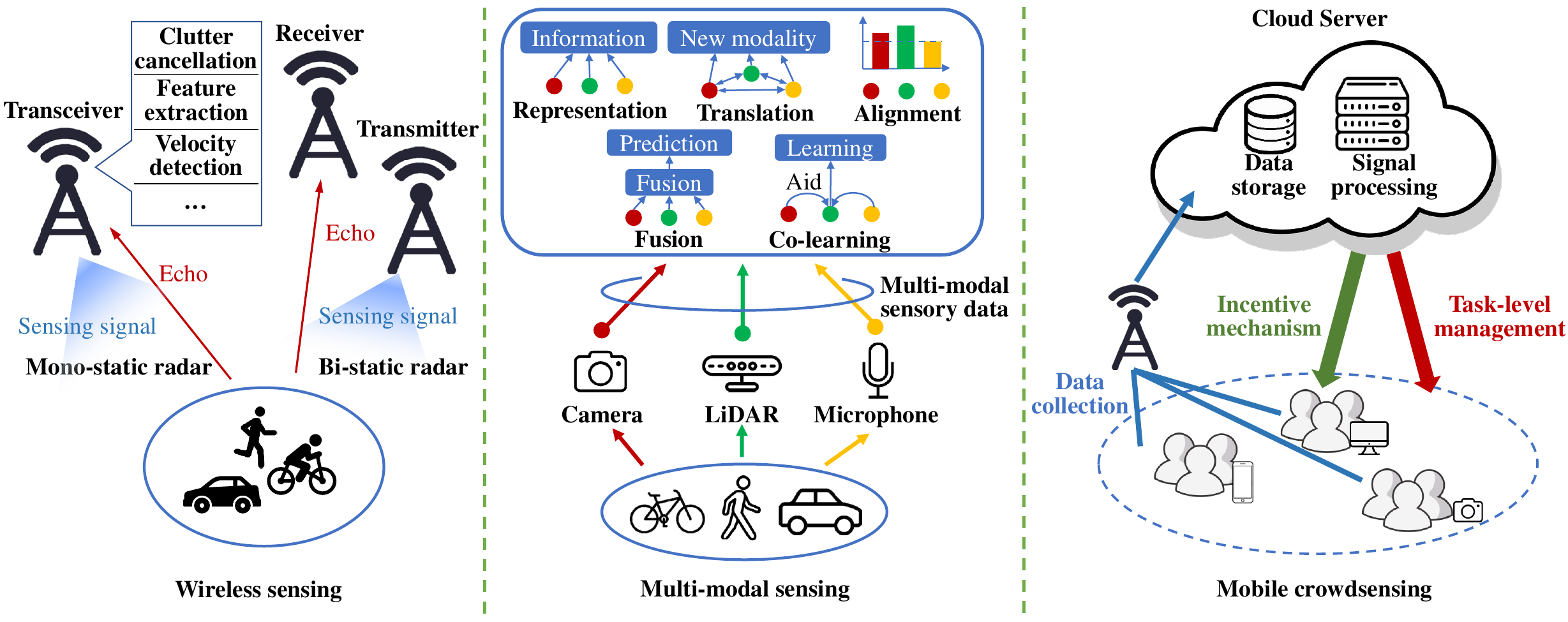}
\caption{Sensing and the corresponding signal processing techniques.}
\label{Fig:SCMP}
\end{figure*}
The first one is wireless sensing (also called radar sensing) technology, which senses a target by receiving and processing the echo signal of a carefully designed waveform. 
Yet, information obtained from a single modality of sensing signals may not be sufficient for completing complicated tasks \cite{suzuki2022survey}. It is mathematically shown in \cite{NEURIPS2021_5aa3405a} that collecting multi-modal sensory data can provide better sensing performance. Therefore, the second ISC technology of multi-modal sensing is developed.  Moreover, although multi-modal sensing technology can enjoy many benefits of rich information, high robustness, cross-modality data transformation, it requires deploying dedicated sensors, which is expensive. To empower ubiquitous sensing in a large coverage with low cost, the third ISC technology called mobile crowdsensing, is proposed \cite{capponi2019survey}. It utilizes built-in sensors of ubiquitous mobile devices for collecting sensory data in various domains. In the sequel, the above three ISC technologies are discussed.

\subsection{Wireless Sensing Technology}

Wireless sensing, which is also called radar sensing, emits modulated radio waves, receives and processes the corresponding echo signals to complete computation tasks such as object ranging and detection, and motion recognition. 
Based on the original continuous-wave radar, new techniques in terms of wave-form design, signal processing, and system architecture are developed to enhance its abilities regarding velocity, distance, direction, and sensitivity toward motions. In this part, the development of radar system design and the corresponding signal processing are introduced. 

\subsubsection{Radar Waveform Design and Signal Processing}
Aiming at achieving higher accuracy of the ranging and detection results, one key advancement of radar techniques is the development of waveform design and the corresponding signal processing approaches \cite{rohling2008radar}. In this part, we will introduce the evolution of the waveform design from the initial continuous wave to the state-of-the-art designs.

\begin{itemize}
\item \emph{Traditional Continuous-wave Radar}:
This type of radar continuously transmits a waveform with a stable frequency and receives an echo signal from one or more targets. 
To be specific, both the transmit signal and the echo signal are sampled with a time interval $T$ and then transformed into the frequency domain using fast Fourier transform (FFT). By calculating the cross-correlation of the two signals and applying a low-pass filter, the Doppler frequency shift (DFS) can be extracted for estimating the velocity. The resolution (or the detection error) of the velocity is proportional to the sampling rate $1/T$. However, the measurement of range information i.e., distance of the target which is decided by light speed and the round-trip delay, is not available with continuous waveform, because the phase shift  of the echo signal is   
ambiguous to the receiver due to the lack of phase basis. That's to say, we cannot distinguish the phase shifts $\theta$ and $(\theta+2k\pi)$ with $k$ being any integer.  

\item \emph{Pulsed Wave Radar}:
To obtain the range information, the pulsed radar is developed, where the transmit waveform is generated by modulating a single-frequency signal by a periodic window signal with period $T$. That's to say, the transmitter periodically produces a brief pulse of radio signal and keeps silent to receive the echo signal of the detected object. The interval between two consecutive pulses is called the inter-pulse period (IPP). The range information is detected by processing the echo signal within one transmission period, called fast-time processing. 
The resolution of rang information is proportional to the pulse transmission period $T$. The velocity is detected using multiple pulses, called slow-time processing. With a fixed velocity, the DFS of each signal in several consecutive pulse waveforms is a single-frequency signal. Hence, by calculating the correlation of the echo signal and the transmit signal over a long time scale (i.e., over multiple transmission periods) in the frequency domain, the DFS is detected and the corresponding velocity can be calculated. The resolution of the detected DFS or velocity is proportional to $1/T$. Note that there is a tradeoff between the resolutions of range and velocity information.  

\item \emph{Frequency Modulated Continuous Wave (FMCW) Radar}:
The time efficiency of pulsed wave radar is relatively low due to the intermittent transmission. To this end, an advanced FMCW radar is developed. The sensing signal is decomposed of multiple repeated snapshots. In each snapshot, a continuous waveform is transmitted with the frequency modulated as a linear function of time. Although transmitting and receiving work simultaneously, no interference is caused as they have different frequencies. Similar to pulsed radar, the range information is detected via fast-time processing, and the velocity is detected by slow-time processing using $M$ snapshots, but the time efficiency of FMCW is higher compared to pulsed radar due to continuous transmission. The resolutions of range and velocity are proportional to the snapshot duration and the number of used snapshots $M$, respectively.
\end{itemize}

\subsubsection{Radar Architectures}
In this part, five radar architectures are introduced.
According to the locations and numbers of transmitters and receivers, radar architectures are categorized into mono-static, bi-static, and multi-static types with each having its pros and cons \cite{hanle1986survey,baker2006bistatic}, which are discussed as follows.
\begin{itemize}
\item \emph{Mono-static Radar}:
The mono-static radar refers to that the transmitter and receiver are co-located \cite{hanle1986survey}.  As a result, this architecture enjoys a concise structure and can be deployed conveniently. More importantly, signal processing is simplified as the synchronization of transmit and receive signals is easy.  However, due to the simple structure, its radar cross section (RCS)  is limited, leading to a relatively poor performance. Besides, continuous waveforms are not feasible in mono-static radar as they suffer from severe self-interference. 

\item \emph{Bi-static Radar}:
In a bi-static radar system, one transmitter and one receiver are separately deployed \cite{baker2006bistatic}. Such systems are featured by the transmitter-target-receiver triangulation. Under this setting, no self-interference is caused at the receiver, leading to a more flexible choice of waveforms. The transmit and receive antennas can be implemented and optimized independently and the RCS-like range and angular resolutions can be improved via spatial filtering. However, the separate design brings more difficult time synchronization between the transmitter and the receiver. Additional information like locations and waveform types should be shared, leading to a higher cost for maintenance.

\item \emph{Multi-static Radar}: 
Multi-static radar systems are the extension of bi-static radar systems \cite{hanle1986survey}. They are consisted of two or more transmitters and receivers with these units separated by large distances when compared to the antenna sizes. Benefiting from this architecture, the probability of target loss is reduced since multiple observations of the target from different aspect angles are enabled. Besides, the energy from a target echo that is created by one transmitter can be utilized by multiple receivers, leading to reduced overall energy consumption for a large sensing coverage as well as less mutual interference. However, this architecture is more complicated, resulting in more expensive costs for time synchronization, information sharing, and joint optimization.

\end{itemize}

To improve radar's sensing ability for object detection in a wider area, early radar systems scan the target via the rotation of a single antenna \cite{Skolnik2022radar}. In this system, an antenna periodically emits a probing signal and receives the reflected signal. As a result, time-division duplex should be carefully designed and waveforms like CW and FMCW cannot be applied. Benefiting from the advancement of multi-antenna techniques, new solutions, i.e., phased array radar and multi-input and multi-output (MIMO) radar as shown in Fig. \ref{Fig:RadarArchitectures}, are proposed for improving the detection coverage of radar, as described below.
\begin{figure}[ht]
\centering
\includegraphics[width=0.48\textwidth]{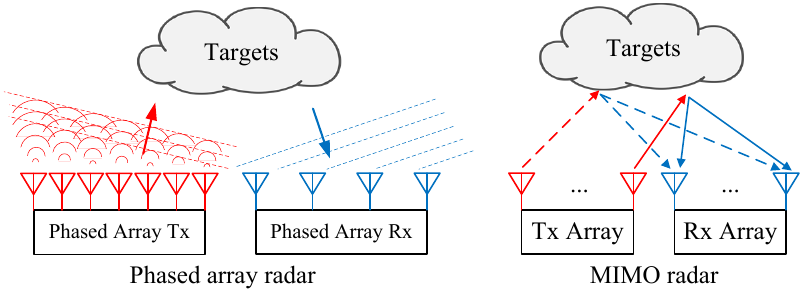}
\caption{Illustration of phased array radar and MIMO radar.}
\label{Fig:RadarArchitectures}
\end{figure}

\begin{itemize}
\item \emph{Phased Array Radar}:  Phased array radar comprises multiple antenna (or radiating) elements arranged in a particular order on a curved or plane surface. Its transmit waveform beam can be electronically steered to points in different directions and distances via controlling the power and phases among these antennas without changing their locations \cite{brennan1961angular,zhang2020sapce}. 
By carefully designing the power and phase control strategy, phased array radar enjoys the advantages of high antenna gain with great side-lobe attenuation, arbitrary directions scanning, high beam agility, multi-functional operations via simultaneous generation of multiple beams, and enhanced robustness, which allows failure of some components. However, the complicated structure that needs to simultaneously control all antennas leads to higher costs and problems like low-frequency agility, deformation of the antenna pattern during beam steering, and limited scanning range.

\item \emph{MIMO Radar}:  In  MIMO radar systems, multiple antennas are equipped at both the transmitter and the receiver \cite{stoica2007on,fishler2006spatial}. In such a system, each transmit antenna radiates arbitrary waveforms independent of other transmit antennas. At the receiver, the echo signal can be mapped with a single transmit antenna due to the independence of waveforms. By paying the cost of higher signal processing complexity, MIMO radar can design more adaptive transmit/receive beamforming and more flexible probing signals to achieve higher detection resolution and better parameter identifiability \cite{robey2004mimo}.
\end{itemize}


\subsubsection{Intelligent Radar Sensing Systems and Applications}\label{Sect:SensingApplications}
Radar sensing shows its superiority compared with other sensing techniques in many aspects, including long-detection distance, high penetration properties, and ease to be equipped in vehicles. As a result, it is widely used in many military applications including military target detection, tracking, and recognition, and civil applications such as disaster assessment, land use and classification, and weather detection \cite{lang2020comprehensive}. However, as the demands of empowering many intelligent services like auto-driving, human detection, and activity classification, modern radar systems encounter new requirements in terms of high accuracy, robustness, and real-time ability. This is beyond the capability of traditional radar signal processing techniques. To tackle this problem, AI, which has shown powerful abilities in many other areas like signal processing in biomedical and communication systems (see, e.g., \cite{subasi2019practical,dahrouj2021overview}), is applied to enable intelligent radar signal processing. According to \cite{9568916}, such techniques can be categorized into five types, i.e., intelligent waveform design, intelligent array design for MIMO radar, radiation source recognition, target detection and recognition, and interference recognition and suppression, which are elaborated below.

Intelligent waveform design refers to using AI models to design waveforms for interference suppression and main lobe gain enhancement of MIMO radar. To suppress the interference of radar signals from other systems (e.g., wireless communication systems) when spectrum sharing is allowed, a reinforcement learning strategy based on deep deterministic policy gradient (DDPG) is adopted to select phases for a phase-coded waveform in \cite{9455187}. A dual Q-learning algorithm with long short-term memory (LSTM) is designed to enable radar signals to use idle sub-bands in \cite{9114698}. On the other hand, to enhance the main lobe gain and minimize the sidelobe radiation in MIMO radar, deep neural networks (DNNs) such as deep residual neural networks \cite{9257002} and feed-forward neural networks \cite{9455163}, are adopted.

Intelligent array design for MIMO radar refers to partitioning the antenna array into multiple subarrays using machine learning methods. The waveforms of different subarrays are designed to be orthogonal in order to achieve a diversity gain \cite{8367469,10548736}. In particular, a convolutional neural network is designed in \cite{elbir2019cognitive} to adaptively select the optimal subarray in a time-varying environment. By training a lightweight convolutional neural network generated by phased-MIMO array manifold matrix, a sparse transmit subarray is selected using a proposed interleaved partitioning method to transmit the reduced-dimensional radar data in \cite{9391717}.

Radiation source recognition identifies and classifies various radar and communication signals based on their emitted waveforms \cite{yang2013hybrid}. Its main goals include detecting and classifying low probability of interception (LPI) radar signals, enhancing the detection accuracy of passive radar systems, and resisting interference or adversarial attacks. Specifically, convolutional and recurrent neural networks are widely used for LPI signal classification \cite{soumya2023recent,10004894}. For improving passive radar accuracy, feature extraction methods like the Wigner Ville Distribution (WVD) are utilized to estimate direct-path signals \cite{10365493}. Moreover, adversarial training and detection strategies are developed to address vulnerabilities to adversarial attacks, which can disrupt signal classification performance \cite{10290894}.


Target detection and recognition involve identifying and classifying objects such as vehicles, ships, and aircraft based on radar signals  \cite{lang2020comprehensive}. In this case, the features such as high-range resolution profiles, Doppler signatures, and synthetic aperture radar images are extracted and fed into different kinds of AI models, including convolutional neural networks, stacked autoencoders, and LSTM networks for fulfilling the tasks \cite{jiang2023radar}. To deal with limited training data, techniques such as generative adversarial networks \cite{10618816} and transfer learning \cite{10246308} are employed to expand the available data set and boost recognition accuracy. Additionally, semi-supervised \cite{10423785} and unsupervised learning \cite{wang2023unsupervised} methods are used when labeled data is scarce, improving recognition performance even with minimal human-labeled input. Moreover, adversarial training techniques are developed to protect the models from adversarial attacks that aim to mislead the classifier, ensuring robust performance in challenging environments \cite{9893068}.


AI has been a powerful tool for jamming/interference recognition and suppression in radar systems. In \cite{10158791}, a fractional Fourier transform based multifeature fusion network uses the local and global features of interference signals and the attention mechanism to improve the interference recognition performance. A multi-scale attention network (MANet) can extract different features through multi-scale dilated convolution, realizing the classification and recognition of target signals and interference signals in ultra-wideband (UWB) signals \cite{10174658}. What's more, by using random signal enhancement and unsupervised domain adaptation to mitigate environmental dependencies in each aspect, a clutter cancellation approach is proposed in \cite{choi2023radar} under the varying environment.

\subsection{Multi-modal Sensing Technology}
A sensory modality refers to the way one entity is observed or perceived, e.g., via radar, camera, microphone, etc. Compared with unimodal sensing like radar sensing, multi-modal sensing can observe or perceive entities through multiple sensing modalities to achieve better task performance. Particularly, the authors in \cite{NEURIPS2021_5aa3405a} mathematically proved that the fusion of data from multiple modalities increases the amount of statistical information and improves the accuracy of learning tasks. Besides, common information can be obtained by different modalities, making multi-modal sensory data more robust to noise and sensory deficiency, i.e., data of one modality can provide complementary or supplementary information to others \cite{8269806}. For example, in an audio-visual speech recognition task, the visual information helps enhance the speech recognition performance when the audio signal is noisy \cite{8269806}. Moreover, the success of the recent deep generative models enables creating one modality of sensory data from another \cite{suzuki2022survey}, such as using text to generate a high-resolution image or a video. This development can bridge the gap between different types of perceptual data, and its applications in computer vision, natural language processing, and other fields are more diverse and creative.

The key design principle for multi-modal sensory data processing is to enhance the performance of downstream tasks by effectively transforming the original data into latent representations suitable for tasks such as object detection \cite{8259006}. This process concerns both efficient encoding and the fusion of common information from diverse sources to maximize the relevance of shared features. The challenge lies in managing the inherent heterogeneity of the data streams, where the variations in quality, format, and temporal alignment must be accounted for to ensure seamless integration \cite{10285741}. The processing procedures of data in different modalities vary and the relation among them is complex. To overcome these challenges and enjoy the previously mentioned benefits of multi-modal sensing,
machine learning algorithms are currently the most powerful and widely adopted methods for multi-modal signal processing \cite{8269806,suzuki2022survey,ramachandram2017deep}. These give rise to a new research area namely multi-modal machine learning. Subsequently, the key issues of multi-modal machine learning are discussed, followed by the introduction of the applications of multi-modal sensing.

\subsubsection{Key Issues in Multi-modal Machine Learning}
As stated in  \cite{8269806,song2024multi}, there are five key issues to be addressed in the area of multi-modal machine learning, namely representation, translation, alignment, fusion, and co-learning. They are elaborated below.
\begin{itemize}
\item \emph{Representation}: Representation or called feature is the information distilled from the data that intelligent systems can use \cite{bengio2013representation}.  
Many efforts have been put into efficiently extracting useful representations for AI tasks based on unimodal data, e.g., representation learning, feature detection, and convolutional neural networks based feature extraction. However, representation extraction from multi-modal sensory data faces new challenges. On one hand, data in each modality is obtained from different types of sensors with heterogeneous forms and noise levels. On the other hand, in many scenarios, data in some modalities is missing. Moreover, it's difficult to measure the correlation and learn a joint distribution of different modalities \cite{8949228}. As pointed out in \cite{srivastava2012multimodal}, a good multi-modal representation should satisfy the following requirements: 1) Similarity in representation should lead to the same similarity in the corresponding results; 2) Multi-modal representation should be readily obtained even in the absence of certain modalities; 3) Missing modalities can be generated using other observed ones. 
The current methods for multi-modal representation can be categorized into two types, i.e., joint representation and coordinated representation  \cite{8269806,song2024multi}. Joint representation projects data of all modalities into the same latent representation space, which is the most adopted method \cite{8269806}. As well as using the conventional neural networks, autoencoders are utilized for pre-training the unlabeled data (e.g., \cite{li2020unicoder})  and probabilistic graphical models like deep Boltzmann machines are adopted to deal with the issue of missing modality via using their generative nature \cite{srivastava2012multimodal}. In coordinated representation, different modalities have their own representations but are coordinated via setting a constraint of similarity (e.g., Euclidean distance) or structure (e.g., partial order) \cite{kuznetsova2012collective}. 

\item \emph{Translation}: Multi-modal translation refers to the ability of modality transfer i.e., generating some modalities of one entity with one source modality \cite{8269806}. How to effectively evaluate the performance of multimodal translation is a problem that needs to be solved. For example, although there is a single answer for tasks such as speech recognition, there is no correct answer for more tasks such as speech synthesis and media description, and there are multiple correct answers for language translation. There are two kinds of translation approaches, i.e., example-based and generative approaches. One is the example-based approach.
In this approach, a dictionary is constructed using the training data samples. For each entity, the source modality and target modality of one entity are stored in the dictionary with one-to-one correspondence  \cite{8269806}. The simplest way called the retrieval-based model for modality translation is finding the closest entry in the dictionary and using it as the translation result \cite{farhadi2010every}. Another way, called the combination-based model, takes one step further. It combines multiple entries (e.g., the most similar $k$ items) in the dictionary with a designated way to create a better translation result \cite{8269806}. 
Although the example-based approach is simple, it calls for expensive costs for the construction and inference of the dictionary especially when the number of items is large  \cite{8269806}. 
The other is the generative approach.
In this approach, generative AI models are trained for modality translation  \cite{8269806}, including two methods. The first is called the grammar-based model, which first extracts high-level concepts from the source modality. Then, these concepts are integrated via a pre-defined grammar-based generation procedure for creating a target modality \cite{barbu2012video}. 
However, this method prefers creating formulaic rather than creative translations, and a separate model and training dataset are required to detect each concept, leading to complex pipelines for concept detection \cite{8269806}. The second method is the encoder-decoder model. It first encodes a source modality into a vectorial representation and then utilizes a decoder to generate the target modality \cite{venugopalan2014translating}. Although this method calls for large quantities of data for training, it is widely used in many generative tasks, like text, image and sound generation, image caption, and video description. 


\item \emph{Alignment}: Multi-modal alignment is defined as finding the relations and correspondences among several modalities of one entity \cite{8269806}. With good alignment, the performance of modality translation can also be enhanced. There are three main issues to be addressed for multi-modal alignment. The first is that not all elements in one modality have correspondences in another, especially in the case of missing modalities. The second is the difficulty in designing a similarity metric for different modalities. The third is the scarce datasets with explicitly annotated alignments \cite{8269806}. To address these issues, two categories of techniques are proposed. 

The first is the explicit alignment method. 
For the case where there are no alignment labels, i.e., labeled correspondences between two modalities, unsupervised explicit alignment algorithms are required. One approach is the linear transformation based approach called dynamic time warping (DTW) \cite{Muller2007}. It measures the similarity between the time sequences of two modalities by time warping and aims at finding an optimal match between them. The alignment performance highly depends on the similarity metric. An extended version of DTW introduces canonical correlation analysis (CCA) to enable both alignment (through DTW) and learning the mapping (through CCA) between the sequences of different modalities \cite{6126545}.
To deal with the non-linear relations, the deep canonical time-warping approach \cite{inproceedings123123}, and various kinds of graphical models based approaches, e.g., generative graphical models \cite{yu2004integration}, factored hidden Markov models \cite{malmaud2015s}, and dynamic Bayesian networks \cite{noulas2011multimodal}, are proposed.
 For the case where there are alignment labels, supervised explicit alignment algorithms are proposed, most of which take inspiration from the unsupervised methods mentioned above, in addition to utilizing the alignment labels to train a better similarity measure (e.g., \cite{plummer2015flickr30k, gebru2017audio}). 

The second is the implicit alignment method. 
It tries to latently align multi-modal data for completing a machine learning task instead of directly aligning the instances of multiple modalities. The early methods adopted graphical models \cite{vogel-etal-1996-hmm}, where the mapping between modalities should be manually constructed. Other methods directly use the neural networks for implicit alignment. Specifically, various kinds of different attention models have been proposed for better performance in question answering and better model interpretability \cite{agrawal-etal-2016-analyzing}, including hierarchical \cite{10.5555/3157096.3157129}, stacked \cite{7780379}, and episodic memory attention \cite{10.5555/3045390.3045643}.


\item \emph{Fusion}: Multi-modal fusion refers to integrating information from multiple modalities of the same phenomenon for completing a prediction task, such as classification and regression \cite{8269806}. As mentioned, using multi-modal information for task completion leads to more robust performance due to their complementary property. However, there are two challenges \cite{8269806}. On one hand, the signals of different modalities may not be temporally synchronized and exhibit different noise levels. On the other hand, it is difficult to build up models that can explicitly exploit the complementary properties of multiple modalities. To address the above challenges, three types of multi-modal fusion approaches are proposed, i.e., early, late, and hybrid fusion \cite{Atrey2010MultimodalFF}. Early fusion refers to integrating the representations or features of multi-modal data and the integrated representations or features are used for completing the downstream task. It can learn the correlation and interactions of different modalities in the early stage and only a single model is trained. Early fusion is widely used in real-time processing, which is useful in scenarios where real-time integration of multi-modal data is required, such as in video surveillance or real-time sentiment analysis combining text and audio. Late fusion is similar to ensemble learning. It fuses the decisions based on unimodal data using mechanisms like averaging \cite{shutova-etal-2016-black}, voting schemes \cite{10.1007/978-3-662-44415-3_16}, or a learned model \cite{conf/acii/GlodekTLSBSKSNPS11, 10.5555/2062850.2062905}. Late fusion uses one predictor for each modality, allowing more flexibility and is more robust to modality missing. The application scope of late fusion is a complex decision-making which is suitable for tasks requiring complex decision-making based on multiple inputs, such as medical diagnosis combining images, patient records, and genetic data. Hybrid fusion, as its name suggests, combines the above two types to enjoy the advantages of both: early fusion for initial comprehensive feature representation and late fusion for refining these representations based on modality-specific processing. Besides,  hybrid fusion methods are more flexible and adaptive to varying data conditions. They can handle scenarios where some modalities might be noisy, incomplete, or missing by allowing for modality-specific processing before integration. This adaptability is crucial in real-world applications where data quality and availability can vary significantly. One of the usage scenarios of hybrid fusion is the adaptive system that needs to adapt dynamically to changing conditions or varying data quality, like in environmental monitoring systems or smart cities.

\item \emph{Co-learning}: Co-learning refers to exploiting knowledge from a (resource-rich) modality to aid the modeling of another (resource-poor) modality when the latter is lack of data, polluted, or has unreliable labels \cite{8269806}. To achieve high co-learning performance, several challenges should be overcome. First, the learned representation of the resource-rich modality may be task-relevant, leading to a limited performance when transferring its knowledge across different modalities in different tasks. Second, the instance numbers and labels are unbalanced. Moreover, the noise levels and the reliability of the labels are heterogeneous. According to whether the multi-modal instances of the same entity are aligned during training, the co-learning approaches are categorized into three types \cite{8269806}. The first is the approch for aligned instances.
In this case, two methods are adopted. One is co-training, where instances of two modalities bootstrap each other with labels for unlabelled instances \cite{8269806}. This method, however, may lead to unbiased training samples and cause overfitting. The other method uses transfer learning-based multi-modal representation learning, which transfers the representation of one modality to another (e.g., \cite{srivastava2012multimodal}).
The second is the approach for unaligned instances.
In this case, the training instances of different modalities are not necessary from the same entities. The first method uses transfer learning to achieve coordinated multimodal representations, e.g., using text messages to improve visual representations for image classification \cite{frome2013devise}. The second is conceptual grounding. It learns semantic meanings or concepts based on data of multiple modalities, including language, vision, and sound, by finding a common latent space of all multi-modal representations (e.g., \cite{baroni2016grounding}).  The last one is zero-shot learning, which recognizes a concept without explicitly seeing any examples of it \cite{8269806}. For example, it can use a neural network to map the text-based message with the extracted visual features.
The third is for the hybrid case, where a pivot modality is used to learn coordinated multi-modal representations using bridge correlational neural networks, or relying on the existence of large datasets from a similar or related task \cite{8269806}.

\end{itemize}

\subsubsection{Applications} 
Multi-modal sensing has empowered various kinds of applications. Via the utilization of audio and visual data, the performance of speech recognition and synthesis  \cite{ma2019unpaired}, human action classification  \cite{yadav2021review}, multimedia event detection \cite{li2020cross}, emotion recognition \cite{kahou2015emonets}, can be greatly enhanced compared with using unimodal information. By utilizing the modality translation techniques, many media description applications have been well developed, including image and video captioning \cite{hodosh2013framing}, visual question-answering \cite{antol2015vqa}, etc. By creating structured representations of multi-modal data, multimedia content indexing, and retrieval applications are enabled, including automatic shot-boundary detection \cite{lienhart1998comparison} and video summarization \cite{evangelopoulos2013multimodal}. Furthermore, multi-modal data can enhance the understanding, analysis, and decision-making processes in the domain of healthcare, fitness, and general wellness \cite{kiros2014stacked,mamoshina2016applications}. Autonomous systems utilize various kinds of sensory data, including radar, stereoscopic visible-light cameras, light detection and ranging (LiDAR), infrared cameras, a global positioning system (GPS), and audio for enhancing the automation capability \cite{maddern20171,geiger2013vision}. 

\subsection{Mobile Crowdsensing Technology}
Mobile crowdsensing, also called crowdsensing or mobile crowdsourcing, has been a powerful paradigm for leveraging built-in sensors of ubiquitous mobile devices like accelerometers, GPS, cameras, and microphones for data sensing and collection in various domains, so as to empower a large number of intelligent and people-centric services \cite{capponi2019survey}. In this sequel, we discuss the advantages, key techniques, incentive mechanisms, and applications of mobile crowdsensing.

\subsubsection{Advantages}
Compared to the traditional sensing paradigm using dedicated sensors, mobile crowdsensing demonstrates its excellence in the following aspects. 
\begin{itemize}
\item \emph{Abundant Data and High Coverage}: The rich set of built-in sensors of massive mobile devices utilized in crowdsensing can collect abundant data and have a much larger sensing coverage \cite{li2018wirelessly}.

\item \emph{Low Cost}: Benefiting from utilizing the existing infrastructure of mobile devices, crowdsensing is more cost-effective in terms of deployment and maintenance \cite{zhou2022energy}.

\item \emph{Scalability}: With billions of mobile devices, crowdsensing is scalable to obtain data from a diverse range of geographical locations, demographic groups, and environmental conditions, making it a promising method for addressing global issues and societal challenges \cite{li2018optimizing}.

\item \emph{Fast Data Collection and Response}: As mobile devices can continuously acquire and transmit data, crowdsensing enjoys the benefits of real-time data collection, immediate analysis, and fast response to emergent and critical missions \cite{zhou2022joint}. 

\item \emph{Innovative and Complex Tasks}: Crowdsensing can empower a wide range of innovative and complex tasks like smart cities, social sensing, and environmental monitoring, via collecting data from multiple resources, different modalities, and various domains (see, e.g., \cite{farkas2015crowdsending,rachuri2010emotionsense,hasenfratz2012participatory}).

\end{itemize}

\subsubsection{Key Techniques}
The efficient deployment of the mobile crowdsesing paradigm depends on the techniques of data collection, data storage, data transmission, signal processing, and task-level management \cite{capponi2019survey}. As the data transmission techniques involve both sensing and communication, we leave its introduction in the next section and the other techniques are introduced as follows. 
\begin{itemize}
\item \emph{Data Collection}: In mobile crowdsensing systems, data collection requires the engagement of massive user participation, where some non-professional users are included. The target is acquiring reliable and trustworthy data via sensor selection. There are two frameworks. One is the domain-specific framework designed for a dedicated task such as intelligent transportation and health care \cite{capponi2019survey}. Its shortage lies in the co-existence of many tasks or domains degrading the efficiency and reliability of the overall crowdsensing system. The other is a general-purpose data collection framework. It supports many tasks simultaneously by maximizing a set of parameters or minimizing the costs \cite{tomasoni2018profiling}. 

\item \emph{Data Storage}: The practical storage location in crowdsensing systems is task-oriented and is classified into centralized storage and distributed storage. The former stores data in a single location, typically a cloud-based database, which is applied when significant processing or data aggregation is needed, e.g., in urban monitoring and price comparison \cite{capponi2019survey}. This approach requires high communication throughput for data transmission and violates the data privacy of users but can enjoy a low processing latency. Distributed storage is employed for delay-tolerant applications such as mapping air quality and fog computing, but has low communication cost and can keep users' data privacy \cite{zhou2022compressive}. 

\item \emph{Signal Processing}: Signal processing in crowdsensing refers to the manipulation and analysis of the collected data. It involves three sequential processes, i.e., pre-processing, analytics, and post-processing \cite{capponi2019survey}. The main target of pre-processing is filtering and denoising by removing irrelevant and redundant information, suppressing noise, aggregating data, and compressing data size to save storage volume \cite{9927216}. Analytics aims to extract meaningful information from the collected data. Machine learning and data mining are common approaches for information inference, pattern recognition, or future trend prediction. For real-time tasks, high computational resources are needed for fast decision-making and response \cite{capponi2019survey}. After analytics, post-processing can be conducted in an offline way for statistical or predictive analysis \cite{6569416}.

\item \emph{Task-level Management}: Since there are a wide variety of crowdsensing tasks, it's significantly desirable to assign and manage the tasks to the participating users or mobile devices. Specifically, task-level techniques include task assignment and scheduling \cite{capponi2019survey}. 

Task assignment involves two aspects. One is task partitioning, which divides complex crowdsensing tasks into smaller and manageable sub-tasks, that can be completed by different devices \cite{wang2016sparse}. For example, in \cite{wei2020sdlsc}, the overall target area is divided into multiple distinct sub-areas, with each being assigned to one device for sensing. The spatial and temporal correlation of different devices' sensory data are leveraged. 
The other task assignment technique is task allocation, which involves distributing tasks to users with the goal of reducing resource costs and enhancing task performance. This technique is categorized into two groups, i.e., centralized methods and distributed methods. In the former, a central coordinator assigns the tasks to users or devices, leading to an enhanced systematic performance but requiring global information (e.g., \cite{wei2020sdlsc}). In the latter, each device makes its own decision during the sensing task, resulting in better data privacy preservation (e.g., \cite{cheung2015distributed}). 

Task scheduling in mobile crowdsensing refers to the way that mobile devices collect and contribute their sensory data. In proactive scheduling, mobile devices actively obtain sensory data without a pre-assigned task and can decide where and when to contribute their data. Proactive scheduling is commonly conducted for people-centric tasks (e.g., \cite{hicks2010andwellness}). In reactive scheduling, the sensing task and target are clearly pre-defined, which is adopted in many works. For example, the authors in \cite{wang2017towards} propose a minimum energy multi-sensor task scheduling algorithm that coordinates multiple tasks for the same user under the targets of enhancing sensory data quality and minimizing energy consumption. 


\end{itemize}

\subsubsection{Incentive Mechanisms}
The crucial condition for the success of mobile crowdsensing is the ability of the network organizer to collect data from the sensors on each user's mobile phone. This, however, violates the data privacy of mobile users and requires additional on-device power consumption for data transmission \cite{gisdakis2016security}. Taking the noise monitoring task as an example, continuous collection of ambient sound violates personal data privacy and reduces devices' battery life. On one hand, many techniques have been proposed to enhance data privacy preservation, including data anonymization by removing personally identifiable information before it is shared \cite{8689081}, data encryption to avoid unauthorized access \cite{9767554}, adding noise under the guideline of differential privacy \cite{jin2016INCEPTION}, and designing secure data-sharing protocols and decentralized data processing methods \cite{xue2019forward}. On the other hand, to encourage as many mobile users as possible to contribute their sensory data with high accuracy and reliability, incentive mechanisms become promising methods to balance the users' additional cost and the overall performance gain of crowdsensing systems \cite{jaimes2015survey, su2020user}. 
Based on whether financial payment is applied, the incentive mechanisms can be categorized into monetary and non-monetary \cite{jaimes2015survey}. In monetary incentives,
paying for sensory data is the most straightforward incentive. 
Some research has shown that the increasing amount of payment can help the sensing task be completed faster (e.g., \cite{mason2009financial}). They are further divided into static mechanisms and dynamic mechanisms. In the former mechanisms, the reward quantity is pre-determined based on a specific criterion, and this amount remains consistent throughout the entire experiment (e.g., \cite{musthag2011exploring}). The latter mechanisms allocate a variable budget for each task, contingent on the real-time conditions of the system (e.g., \cite{nan2014cross}). 
Non-monetary incentives have no financial payment and call for voluntary participation \cite{ben2011galaxy}. This severely restricts its scope of application. In some specific tasks, contributing users can benefit from the overall crowdsensing tasks, making it possible to design appropriate incentive mechanisms, including collective motives, social reward, and intrinsic motives and fun \cite{jaimes2015survey}. Collective incentives encourage the achievement of common goals, which is the philosophy of crowdsensing, to work together for a common good. However, once the goal is reached, non-participants can also make a profit. Hence, it's desirable to add additional incentives to avoid  ``a free ride". 
In non-monetary incentives, the inherent interest and personal fun, like gaming, are the motivation of the majority of participants \cite{wiggins2011ebirding}. 


\subsubsection{Applications}
By leveraging the rich data collected via mobile crowdsensing, a wide range of applications are empowered, including emergency management and prevention for accidents and natural disasters \cite{ludwig2015crowdmonitor}, environmental monitoring \& waste management applications such as air quality \cite{hasenfratz2012participatory}, noise pollution \cite{schweizer2011noisemap}, and waste-recycling operations \cite{cuff2008urban}, E-commerce applications for collecting, processing, and delivering prices of goods in the real physical world \cite{sehgal2008mobishop}, mobile social networks that build social relations and meet and share data between people with similar interests using mobile devices \cite{chon2013understanding}, public safety applications for evaluating the safety exploiting the urban environments \cite{kantarci2014trustworthy}, health care and wellbeing applications \cite{bort2014measuring}, intelligent transportation systems \cite{habibzadeh2018sensing}, etc.

\subsection{Discussion}

All of the above three ISC technologies have vast applications, especially in the era of AI. Each of them has its own pros and cons. Wireless sensing enjoys the benefits of simple data format, mature signal processing techniques, long-detection distance, high penetration properties, and ease of being equipped in vehicles. The key technical issues of wireless sensing include architecture design, waveform design, and signal processing to overcome the challenges caused by the hostile wireless environments. Multi-modal sensing is more robust and can enhance task performance by exploiting and fusing the information from data in multiple modalities as already mentioned. However, the key issues like representation, alignment, and fusion, call for complicated AI-based algorithms. Besides, there is a lack of theoretical methods to guide its design.  Mobile crowdsensing has the advantages of abundant data, high coverage and scalability, low cost, and fast response.  However, it takes the risk of violating data privacy and thus requires more efforts to encourage users to contribute data.

\section{Integrated Sensing and Communication}

\subsection{Overview}

ISAC has been a promising technique in wireless networks for saving network resources and enhancing the performance of both sensing and communication. Different from existing literature (e.g., \cite{liu2020joint,paul2016survey,hernandez2022wifi}) focusing on integrated wireless sensing and communication, a comprehensive survey about ISAC is provided in this work. Specifically, three ISAC paradigms will be presented which align the three ISC technologies in Section \ref{Sect:ISC}, as shown in \ref{Sect:ISC}  and listed below.
\begin{itemize}
\item \emph{Integrated Wireless Sensing and Communication}:  This paradigm enables hardware and radio resources sharing between wireless sensing and communication. The general integrated wireless (radar) sensing and communication technique is first elaborated, including two frameworks of radar communication coexistence (RCC) and dual-functional radar-communication (DFRC) \cite{paul2016survey}. Then, a special case of Wi-Fi sensing is discussed, since WiFi signals are widely distributed and can be used for sensing at low cost. 

\item \emph{Integrated Multi-modal Sensing and Communication}: Utilization of only uni-modal electromagnetic wave for ISAC cannot provide complete and fine-grained characteristics of both sensing target and wireless channels \cite{cheng2024intelligent}. This motivates the emergence of integrated multi-modal sensing and communication. In this work, three key techniques are introduced, i.e., dataset construction, multi-modal sensing for channel estimation, and communication for multi-modal sensing enhancement.

\item \emph{Integrated Mobile Crowdsensing and Communication}: The current research focus on this topic lies in using existing communication techniques for collecting massive distributed sensory data to a central server for completing the downstream tasks. The integration of crowdsensing and communication from a systematic view remains an unexplored area.  
\end{itemize}

\begin{figure*}[ht]
\centering
\includegraphics[width=1\textwidth]{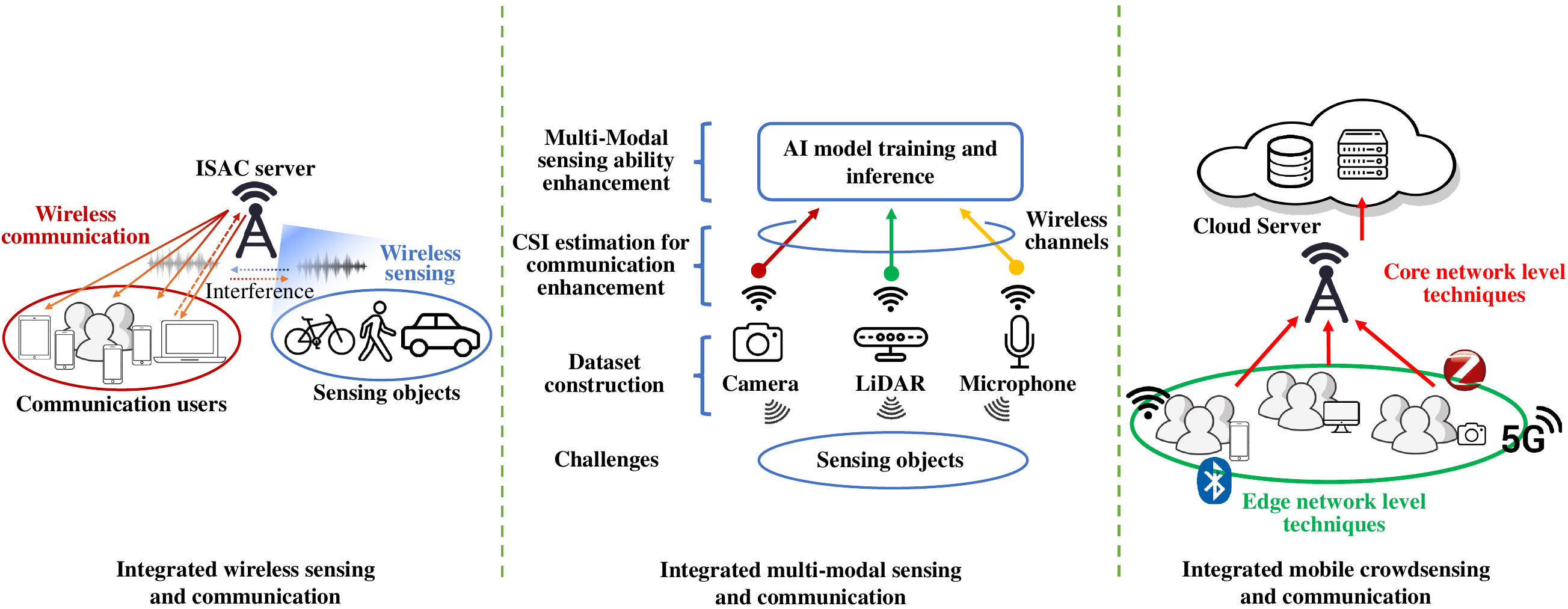}
\caption{Illustration of three ISAC paradigms.}
\label{Fig:SCOM}
\end{figure*}


\subsection{Integrated Wireless Sensing and Communication}\label{Sect:IWSC}
\subsubsection{General Integrated Wireless Sensing and Communication}
With the explosion of mobile traffic, the available frequency bands for communication are crowded, which drives an urgent demand for more spectrum resources. One promising solution is to reuse the wide radar spectrum \cite{griffiths2014radar}, giving rise to an emerging research topic i.e., integrated wireless sensing and communication \cite{liu2020joint}. On the other hand, many radar sensors are deployed for significant applications like air traffic control, geophysical monitoring, surveillance for defense, etc. The rising interference from wireless communication systems should be well managed and suppressed. To tackle these issues caused by communication and radar spectrum sharing, two frameworks are widely investigated, namely, RCC and  DFRC \cite{paul2016survey}. The former aims to develop efficient interference-suppression techniques so that the two systems can coexist without interference. The latter focuses on designing systems that can simultaneously realize wireless communication and remote sensing. In the sequel, the techniques for empowering these two frameworks are elaborated.

The main issue in an RCC system is to avoid interference between radar and communication. Recent research starts from designing opportunistic spectrum access and interference management or cancellation schemes based on interference channel state information (ICSI) estimation.  Similar to cognitive radio, radar is regarded as the primary user and the communication system plays the role of the secondary user. 
To be specific, a communication transceiver intelligently detects whether a communication channel is in use or not and moves transmission into vacant channels. This makes better use of spectrum resources and enhances the frequency efficiency. To avoid the possible performance degradation of radar caused by co-channel interference, the communication system should control its transmit power \cite{wang2008application}.   Although it is simple to implement opportunistic spectrum access schemes with low cost, there is no real spectrum sharing between radar and communication. Besides, such schemes cannot facilitate the coexistence of communication and MIMO radar, since it's difficult for the former to identify the sidelobes with radar's omnidirectional waveforms. Therefore, the most recent work focuses on investigating frequency-sharing schemes under the target of interference cancellation based on ICSI estimation. 

The key to support spectrum sharing between radar and communication is ICSI estimation. Conventionally, ICSI is estimated by exploiting pilot
signals sent to the radar by the communication system \cite{mahal2017spectral}. An alternative method is to set a dedicated coordinator connected to both systems via wireless or backhaul links for ICSI estimation and transmit precoding design \cite{li2017joint}. These methods, however, cause additional computational and signaling costs. To alleviate the estimation overhead, the radar probing waveform can be utilized as the pilot signal \cite{liu2019interfering}. Based on the knowledge of ICSI, different techniques are proposed for supporting the spectrum sharing between radar and communication systems, as discussed below.

\begin{itemize}
\item \emph{Null-space Projection (NSP) Radar Precoders}: In this scheme, the radar precoder is designed to transmit waveforms only in the null space of the interference channel matrix, which is obtained by applying singular value decomposition (SVD) \cite{khawar2015target}. As a result, the interference from radar to communication is strictly avoided. However, this scheme limits the performance of radar, especially for MIMO radar where the null space of the interference channel is small. To address this issue, some other schemes enhance the radar performance by allowing non-zero interference in the dimensions with small singular values at the communication receiver (see, e.g., \cite{babaei2013nullspace}).  

\item \emph{Receiver Design}:
Most works on this topic focus on accurately demodulating the communication data by suppressing the radar interference. By exploiting the sparse property of the radar interference and the communication error caused by channel fading and noise,  compressive sensing is utilized to recover the radar interference and the communication signal simultaneously \cite{zheng2017adaptive}. For radar using pulsed waveforms, its interference either has high amplitudes in short durations or is zero otherwise, which keeps unchanged for a relatively long time scale due to the slow variation of radar settings. Thus, the interference amplitude can be accurately estimated with low cost. Benefiting from this fact and by assuming the phase shift of radar interference in a uniform distribution in the range of $[0, 2\pi]$, the optimal demodulating scheme is first investigated under given modulation constellations, followed by the design of self-adaptive constellations that optimize communication capacity and the symbol error rate \cite{nartasilpa2018communications}.

\item \emph{Joint Design}: 
In this case, the radar and communication system parameters such as transmit power and beamforming are jointly adjusted. The pioneering work considers the coexistence of a matrix completion MIMO (MC-MIMO) radar and a point-to-point MIMO communication system \cite{li2016optimum}. It randomly sub-samples the radar receive antenna array to enlarge the null space, which provides more degrees of freedom for the communication system without causing interference to the radar system. Particularly, the communication signal's covariance matrix and the radar's antenna down-sampling matrix are jointly optimized. To address the case with signal-dependent clutter, the authors in \cite{li2017joint} optimize the radar signal-to-interference-plus-noise ratio (SINR) while guaranteeing communication quality. Furthermore, the joint optimization scheme is proposed in \cite{liu2017robust} under the assumption of imperfect ICSI. 

\end{itemize}

In a DFRC system, radar and communication can be simultaneously performed and jointly designed via sharing hardware platforms and spectrum resources. However, different from communication systems, radar systems have uncooperative targets and the useful information is in the echo signals \cite{bliss2014cooperative}. This leads to different design principles. To unify the design of radar and communication, researchers have made efforts to investigate the performance metrics of radar from an information-theoretic view. In \cite{guerci2015joint}, the capacity of a round-trip radar channel is defined as the number of distinguishable targets.  Furthermore, the authors in \cite{chiriyath2015inner} point out that the maximum achievable information of detected targets of a radar system is the mutual information of the transmit signal and the echo waveform. Thereby, the target of echo signal processing is to reduce the uncertainty of detected targets for enhancing the detection precision. The remaining part of the uncertainty is lower-bounded by the CRB \cite{kay1993fundamentals}. On this basis, the performance tradeoff between communication and radar is investigated \cite{chiriyath2015inner,chiriyath2017radar,rong2018mimo}. 

Another main research topic of DFRC is to design dual-functional waveforms useful for both radar and communication. The pioneering work utilized pulsed interval modulation to modulate communication bits on the radar pulsed waveforms \cite{mealey1963method}. On top of this method, many work designed dual-functional waveforms for both communication and radar via different modulation schemes (see, e.g., \cite{roberton2003integrated, saddik2007ultra, jamil2008integrated}). Whereas, a better approach is to exploit the existing communication signals for completing sensing tasks. Particularly, the widely used OFDM signals in modern communication systems are regarded as a promising candidate \cite{garmatyuk2010multifunctional}. Fast FFT and IFFT are utilized for estimating the DFS and range information \cite{sturm2011waveform} from the OFDM waveforms. What's more, researchers further target designing dual-functional waveforms in MIMO systems via beamforming. One basic approach is to use the main lobe of the MIMO signal for target detection and transmit information via the sidelobe \cite{hassanien2016non,boudaher2016towards}. However, in this approach, communication symbols are embedded into the radar pulses, and thus the communication data rate depends on the pulse repetition frequency (PRF) of the radar, which is much lower than the conventional communication systems. Besides, communication is not available when the sidelobe has no line-of-sight link to the receiver. To address these shortages, better schemes directly use the communication waveforms for target detection, which optimize the transmit beamforming for balancing the performance tradeoff between radar and communication (see, e.g., \cite{liu2018mu,liu2018toward}). Moreover, a series of tecnologies are exploited to improve the performance of ISAC, such as the intelligent reflecting surface (IRS) \cite{guo2023joint} and wireless power transfer \cite{li2023multi,li2022wirelessly,li2023integrating,zhou2023integrating}.

\subsubsection{Wi-Fi Sensing}
As its name suggests, Wi-Fi sensing, as a special case of integrated wireless sensing and communication, exploits Wi-Fi signals for detection, recognition, and estimation \cite{hernandez2022wifi}. Compared with traditional radar sensing, it has a lower cost since the Wi-Fi signals and equipments are ubiquitous and no extra hardware is needed \cite{chen2024deep}. Wi-Fi sensing can be widely applied in a series of scenarios such as fall detection \cite{cai2023falldewideo}, indoor tracking \cite{wang2023ukfwitr}, and so on. Particularly, OFDM signals are used for sensing, since OFDM modulation is widely adopted for efficient communication in Wi-Fi systems by dividing the frequency-selective broadband channels into multiple orthogonal subcarriers with flat fading in each subcarrier \cite{hernandez2022wifi}. The essential data for completing Wi-Fi sensing tasks is CSI of OFDM signals, as CSI can describe the target effects on amplitude and phase of the signal and can provide rich information over multiple subcarriers \cite{hernandez2022wifi}. The collected CSI is first processed before inputted into AI models for finishing the tasks, i.e., detection, recognition, or estimation. In the sequel, the signal processing techniques are described.

The first important signal processing technique is to extract useful features from the raw CSI data. The extracted features include the amplitude and phase of signals, their difference over subsequent time slots to capture the temporal changes as features \cite{zhou2023towards}, the statistical features like mean, standard deviation, median, kurtosis (see, e.g., \cite{zhang2023ratiofi}), the signals' power density in the frequency domain (see, e.g., \cite{damodaran2020device}), or the time-frequency components generated by wavelet transform  (see, e.g., \cite{wang2020csi}). Then, it's significant to cancel the interference and noise in the received Wi-Fi signals before being used for completing the sensing task, since interference from other sources and reflected by the environmental obstacles is very common in Wi-Fi systems \cite{hernandez2022wifi}. To cancel the interference, many typical filters are used, including moving average over several time slots (see, e.g., \cite{huang2021wilay}), Hampel filter \cite{zuo2021new}, and FFT frequency filter \cite{tan2016wifinger}. Subsequently, the collected CSI sample often has redundant information since signals over adjacent subcarriers have similar contents \cite{tan2020enabling}. To exclude the redundant information for reducing computation overhead of downstream tasks (e.g., detection and recognition), dimensionality reduction is performed, like selecting subcarriers with large temporal variance (see, e.g., \cite{tan2020enabling}), deleting the subcarriers with negative correlation to others (see, e.g., \cite{zhuang2021develop}), using PCA to select the principal eigenspace of all subcarriers (see, e.g., \cite{wang2015understanding}), etc. 

In addition to the common signal processing techniques, i.e., feature extraction, interference cancellation, and dimensionality reduction, mentioned above, there are many application-specific pre-processing approaches.  Specifically, the method of detrending targets mitigates the influence of fluctuating trends caused by environmental change via using approaches like least square regression \cite{yu2021wifi}.  Interpolation is applied when there is packet loss in the collected CSI \cite{zhao2024finding}. Segmentation is exploited to decide how many CSI samples should be collected over time for the following decision-making using e.g., a window with a specific start and stop sentinel movement \cite{li2020novel}.


\subsection{Integrated Multi-Modal Sensing and Communication}
Although integrated wireless sensing and communication have achieved significant development, such techniques cannot fully utilize the massive non-RF sensors and the corresponding sensory data \cite{cheng2024intelligent}. What's more, in future wireless networks, the electromagnetic environment may change rapidly, leading to severe frequency selectivity and time selectivity in wireless channels. The performance gain from integrated wireless sensing and communication, however, comes from the electromagnetic wave propagation domain, which cannot provide either complete or fine-grained characterization of the rapidly changing environment. This leads to little performance enhancement potential \cite{cheng2024intelligent}. On the contrary, multi-modal sensing systems obtain information of the environment through multiple domains, including radio waves, visible light, or infrared light reflections. With the widely deployed multi-modal sensors, the parameters of the physical environment can be acquired more precisely and efficiently, thereby enhancing communication efficiency. On the other hand, with fast and reliable communication networks, multi-modal sensory data can be shared, exchanged, and fused among different devices, which can further significantly enhance the sensing quality and the performance of downstream tasks such as AI model training and inference. To this end, the research paradigm of integrated multi-modal sensing and communication is proposed \cite{cheng2024intelligent}. In the sequel, the challenges and techniques are first elaborated. Then, a promising research topic of integrated multi-modal sensing and communication called haptic communications, is introduced \cite{antonakoglou2018toward}, which is regarded as a key technology in 6G.

\subsubsection{Challenges and Techniques}
The efficient implementation of integrated multi-modal sensing and communication faces two main kinds of challenges \cite{cheng2024intelligent}. On one hand, the obtained multi-modal sensory data are in various formats and differ in physical properties like resolution, bandwidth, and latency. To characterize the rapidly changing wireless channel information, it calls for fast and accurate data collection and processing methods.
On the other hand, there is currently a lack of datasets that contain aligned multi-modal sensory data and wireless channel, especially for customized scenarios, due to labor and cost concerns. As a result, the mapping relation between multi-modal sensing and communication remains an unexplored issue. 

The techniques of integrated multi-modal sensing and communication can be grouped into three types. The first type targets constructing datasets for building up the relation between multi-modal sensory data and communication environments. The second type aims to exploit the multi-modal sensory data and machine learning algorithms for CSI estimation. The last type enhances the performance of multi-modal-sensing-based intelligent tasks. 
\begin{itemize}
\item \emph{Dataset Construction}: As mentioned, the multi-modal sensory data collected from different types sensors have heterogeneous formats and physical properties which results in a lack of processing methods with interpretability and expansibility. Besides, the mapping relation between multi-modal sensing and communication is unexplored. To tackle these issues, a good dataset of integrated multi-modal sensing and communication should satisfy the following requirement \cite{cheng2024intelligent}. To begin with, the sensory data and wireless channel data should be aligned to achieve mutual promotion between each other. Then, the sensory data should be collected from various devices in order to enlarge the sensing ranges, thereby enhancing the scalability and generality of the supported applications. Moreover, the dataset should be obtained from real scenarios to make sure that its supported research and projects have practical and high-generation ability. To this end,  a vision-aided wireless communication dataset is constructed using  3D modeling and ray-tracing simulators \cite{alrabeiah2020viwi}, where  RGB images and depth maps are created from the visual instances, and the wireless features are obtained with sinusoid waveform generator. By exploiting the dataset in \cite{alrabeiah2020viwi}, a machine-learning-based framework is proposed in \cite{2102-09527}  for enabling proaction in high-frequency wireless networks.  The authors in \cite{cheng2023m3sc} develop a simulation dataset called M$^3$SC in dynamic vehicular communication networks, where the multi-modal sensory data and wireless channel data are aligned and massive MIMO and millimeter wave (mmWave) communications are considered. For better characterizing the practical imperfections, a large-scale integrated multi-modal sensing and communication dataset is proposed in  \cite{2211-09769}, where the data are collected from real-world scenarios with mmWave receivers, RGB cameras, 3D LiDARs, RF radars and GPS receivers. 

\item \emph{CSI Estimation for Communication Enhancement}: Multi-modal sensory data include information on key elements in the environments such as the position of scatters, that are highly relevant to CSI. Therefore, using multi-modal sensory data for CSI estimation has attracted much focus. However, the mapping relation between multi-modal sensory data and communication characteristics is difficult to obtain. Many works manage to build up such a relation via using DNNs (e.g., \cite{8052521,9665355,9900133}). Another research direction targets enhancing the communication performance in MIMO systems, where beam should be aligned to increase the received SNR. Multi-modal sensory data has been useful for predictive beamforming, where the position of a target is predicted with the assistance of multi-modal sensors so as to perform the corresponding beamforming in advance (e.g., \cite{9923616, 9771564}). Particularly, the use of multi-modal data for position prediction enjoys the benefit of higher robustness and accuracy due to the common and supplementary information provided by data in different modalities.

\item \emph{Multi-Modal Sensing Ability Enhancement}: In multi-agent multi-modal sensing systems, communication plays an important role in data fusion among different devices for e.g., obtaining beyond-vision global and fine-grained sensory data. There are three cross-device multi-modal data fusion schemes. The first is raw data fusion, where raw sensory data is shared among different devices, leading to the largest information redundancy and communication overhead \cite{cheng2024intelligent,6083061, 6866903, 8885377}. To overcome the communication bottleneck in raw data fusion, the feature fusion scheme is proposed, where low dimensional feature vectors are extracted and shared at each device  \cite{10.1145/3318216.3363300, Marvasti2020FeatureSA,9330564, 9718512}. The last scheme is semantic fusion, which requires the least communication bandwidth. In this scheme, the semantic results of different devices are shared \cite{8551565, 8569832}.

\end{itemize}

\subsubsection{Haptic Communications}
Haptic communications refer to exchanging haptic information in networked teleoperation systems for supporting remote immersive applications, including tele-diagnosis, tele-surgery, remote teaching, virtual/augmented reality, and so on \cite{antonakoglou2018toward}. Haptics include multiple modalities of sensory data from both kinesthetic perception (including forces, torques, position, velocity, etc.) and tactile perception (including surface texture, friction, etc.) \cite{antonakoglou2018toward}. Although traditional remote interactions such as voice or video can well support many remote applications like remote voice and video conferencing, the achievement of remote immersion demands additional haptic information \cite{antonakoglou2018toward}. Last but not least, the requirements of high packet rate, low packet loss, and variable delay lead to hard management and synchronization of the data streams \cite{nitsch2012meta}. 

The challenges of haptic communications lie in three aspects. The first is the huge communication load caused by the haptic sensors. To guarantee the stability and user experience of teleoperation systems, many types and quantities of haptic sensors should be deployed and they obtain and transmit data at a high sampling rate, e.g., exceeding 1 kHz  \cite{tan1994human}, which brings huge communication overhead.  Besides, remote immersive applications, especially critical missions such as telesurgery, are very sensitive to data loss and latency \cite{lawrence1993stability}. In summary, haptic communications demand ultra-high reliability and low latency \cite{chen2017green}.

Existing techniques focus on improving the communication efficiency for transmitting the haptic sensory data, generally together with the corresponding video and audio data. The correlation among the signals of different modalities can be utilized for data compression thus alleviating communication overhead \cite{wei2021cross}.  A cross-modal communication architecture is proposed in \cite{chen2023toward}, where the heterogeneous modal requirements are satisfied and the cross-modal correlation is utilized for data reconstruction. Besides, data in different modalities has different transmission rate demands \cite{zhou2019cross}. For example, the transmission of audio/video data calls for high throughput, whereas that of haptic data demands ultra-high reliability and low latency \cite{chen2017green}. Based on this property, several modality-aware communication techniques are designed via e.g., setting priority for haptic data \cite{cizmeci2017multiplexing} and {adaptive} resource allocation among different modalities \cite{gao2022edge}.

\subsection{Integrated Mobile Crowdsensing and Communication}

Typical mobile crowdsensing services are implemented on central servers or clouds. However, the collection of a huge amount of sensory data from massive mobile devices causes a large communication overhead. Besides, most mobile devices communicate via hostile wireless links with channel fading, noise, and limited network resources. Furthermore, many crowdsensing services demand real-time decisions and fast responses, resulting in more severe communication bottlenecks. To maximize the overall crowdsensing system utility calls for communication techniques to overcome the above drawbacks. A direct approach adopts the existing communication schemes in telecommunication networks. They are divided into the edge-network-level techniques and the core-network-level techniques \cite{capponi2019survey}. The former collects sensory data in local wireless edge networks via utilizing existing protocols, including  ZigBee, Bluetooth, WiFi, and cellular communications. The latter targets remote data transmission e.g., from mobile devices to clouds, by using a high-speed fiber-optic architecture to provide connections among different wireless edge networks and clouds. Although directly adopting existing communication schemes has low design complexity and is cost-effective, the separated design between communication and crowdsensing degrades the resource utilization and cannot fully unleash the performance potential \cite{antonic2016mobile}. To this end, a joint design that integrates crowdsensing and communication from a systematic view is desirable. There is currently little work focusing on this area, leaving many research opportunities. One possible research direction lies in adaptively designing the data sensing, collection, and processing approaches according to the time-varying wireless channels and communication capacities. Another potential research direction is joint resource allocation among the processes of sensing, communication, data filtering, and quantization to maximize the data utility.

\subsection{Discussion}
The three ISAC paradigms have their own advantages and challenges. The paradigm of integrated wireless sensing and communication enjoys the benefit of low cost since it can reuse the ubiquitous signals and hardware in wireless communication networks for sensing. Besides, the techniques of wireless signal processing and transceiver design are mature. Moreover, the existing methods in communication systems, such as MIMO beamforming and OFDM modulation, can be directly implemented in this paradigm. These make it easy to implement. However, the interference management between sensing and communication as well as the requirements to achieve both goals of the two tasks bring new design challenges. In the paradigm of integrated multi-modal sensing and communication, communication enables cooperative multi-modal sensing among many sensors to complete complicated tasks and achieve higher task performance. On the other hand, multi-modal sensing can provide abundant prior knowledge for enhancing communication quality. The challenges lie in two aspects. One is that the multi-modal sensory data are in various formats and differ in physical properties like resolution, bandwidth, and latency. The other is the lack of alignment, labeling, and fusion techniques for constructing practical datasets. For integrated mobile crowdsensing and communication, it enjoys the benefit of collecting massive data for completing many tasks to improve the data utility. The challenges arise from the data privacy, the huge communication load, and the characterization of the data utility function.

\section{Benefits, Functions, and Challenges of ISCC}

ISCC refers to the integrated design of sensing, communication, and computation modules via resources (e.g., energy, time, spectrum, and hardware) sharing and coordination, and joint algorithm development among them. In thi section, the benefits of ISCC, the functions of ISCC in 6G, and its challenges are discussed.


\subsection{Benefits}

Compared with existing partial integration technologies, ISCC enjoys the benefits of {\bf sensing-communication-computation symbiosis} and overcoming the three shortages mentioned in Section I-B to achieve {\bf higher resource utilization and task performance}.

\begin{figure}[h]
\centering
\includegraphics[width=0.48\textwidth]{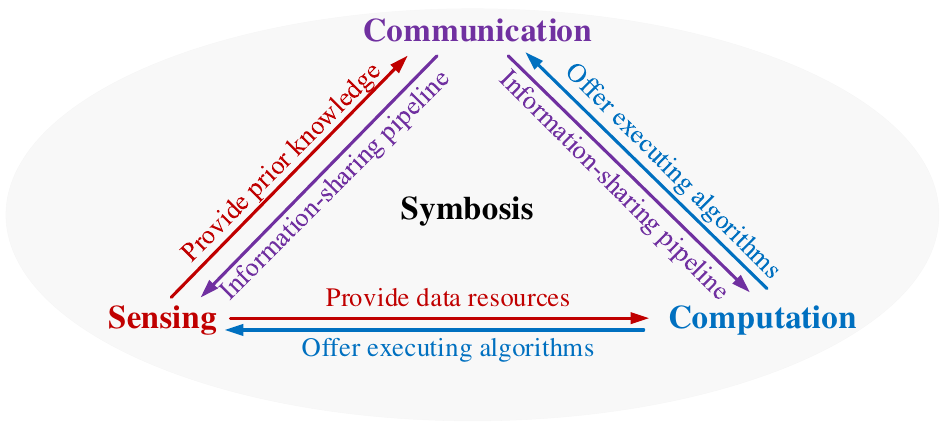}
\caption{Sensing-Communication-Computation Symbiosis.}
\label{Fig:Symbiosis}
\end{figure}
To begin with, ISCC has the opportunity for sensing, communication, and computation symbiosis, as shown in Fig. \ref{Fig:Symbiosis}. Specifically, the information obtained via sensing can be used as prior knowledge to enhance communication quality and the data resources for executing complicated computational tasks, e.g., federated learning and edge AI inference. Communication serves as an information-sharing pipeline for cooperative sensing and distributed computational tasks. Computation can provide intelligent and efficient execution of algorithms for enhancing the sensing and communication performance. This ISCC-enabled closed-loop framework can achieve mutual performance improvement in all three modules. 

Then, ISCC can achieve higher resource utilization than the three partial integration technologies. In the networks that are enabled with partial integration technologies, network resources are partially shared between two modules or even orthogonally shared among the three modules. For example, in ISAC systems, radio and hardware resource sharing is enabled between communication and sensing tasks, but not for computation tasks \cite{cui2021integrating}. However, in an ISCC system, the waveforms corresponding to the sensing communication, and computation tasks can coexist in the same radio resource block \cite{qi2020integrated} or one waveform is directly designed for completing the three kinds of tasks \cite{li2023integrated}. On the other hand, compared with existing partial integration technologies, where network resources are partially coordinated between one or two modules, network resource management among sensing, communication, and computation modules from a systematic view is enabled. For example, in ISCC-enabled edge-device cooperative systems, ISCC resource allocation can save resource costs to achieve the same performance compared with separate resource allocation schemes \cite{wen2024integrated,zhuang2024integrated}.

\begin{figure}[h]
\centering
\includegraphics[width=0.5\textwidth]{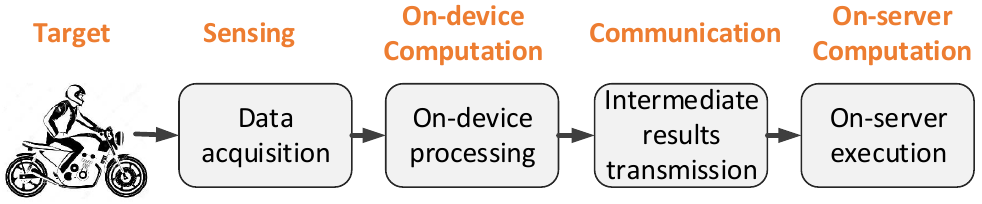}
\caption{The sensing, computation, and communication processes in complicated tasks.}
\label{Fig:SCC_Processes}
\end{figure}
Moreover, for complicated tasks that involve sensing, computation, and communication processes as shown in Fig. \ref{Fig:SCC_Processes}, existing partial integration technologies face shortages of mismatch between the goals of the overall task and each module and ignoring the tight coupling of different modules, as mentioned in Section \ref{Sect:ISCC_Motivation}. The former shortage refers to the design metrics of each module in existing partial integration technologies, e.g., mutual information for sensing, throughput for communication, and MSE for AirComp, don’t match the goal of the overall complicated task. As an example, in an edge AI inference task, enhancing the sensing quality of the least important view or transmitting the irrelevant feature elements may enhance the traditional design goals of sensing and communication but is useless for improving the inference accuracy. The latter shortage refers to existing partial integration technologies failing to respect the coupling mechanism where the task performance depends on the quality of each module but different modules compete for network resources. For example, in an ISCC-supported online federated learning framework proposed in \cite{wen2024AirCompISCC}, the convergence rate is decided by the number of sensory data samples, the sensing SNR, the communication SNR, and the computation speed, but the three modules demand to compete for time and energy to improve their own higher qualities. ISCC can overcome the above two shortages to achieve higher task performance by adaptively sharing, coordinating, and allocating network resources among sensing, communication, and computation in a task-oriented way, i.e., under a unified task goal.

\subsection{ISCC's Functions in 6G}\label{Sect:6GFunctions}

ISCC will be a key enabler to enhance system performance in 6G and beyond networks. In the sequel, we shall discuss how ISCC can enhance 6G KPIs and facilitate the usage scenarios.

\subsubsection{ISCC Enhances 6G KPIs}

With the advantages of sensing-communication-computation symbiosis, higher resource utilization, and better task performance, ISCC can help achieve many 6G KPIs. Some examples are listed below.
\begin{itemize}

\item \emph{Spectrum Efficiency}: ISCC enables different kinds of waveforms (i.e., sensing, communication, and computation) to share the same frequency band \cite{qi2022integrating} or directly utilize one waveform to complete the three functions \cite{li2022inte}. Besides, in ISCC systems, AI methods can be utilized to process real-time sensory data for spectrum detection and prediction \cite{10627103}. As a result, the spectrum efficiency is improved.

\item \emph{Mobility}: By carefully designing the multi-modal sensing strategy toward moving communication devices, their velocity, positions, and trajectories can be estimated or predicted. This can help compensate for the Doppler effect and enable soft handoff among different cells, especially for fast-moving devices. 

\item \emph{Task-completion Latency}: By adaptively allocating the time among sensing, communication, and computation modules and allowing pipeline design for long-run tasks, e.g., allowing time overlapping between sensing for obtaining a new data sample and the extraction and the transmission of the local feature vectors in continuous AI inference tasks, the end-to-end task completion latency is saved.

\item \emph{Connection Density}: In a dense network, the devices' locations and channel gains can be estimated by intelligently processing the real-time multi-modal sensory data. Therefore, better beamforming, time, or subcarrier allocation strategies can be designed to enhance the connectivity. 

\item \emph{Energy Efficiency}: ISCC can help enhance the network energy efficiency. On one hand, it allows reusing signals from one transmitter for completing multiple tasks of sensing, communication, and computation. On the other hand, the energy of devices and servers can be adaptively managed among the three processes. 

\end{itemize}

\subsubsection{ISCC for 6G Usage Scenarios}
As a key technology in 6G, ISCC is expected to support a series of usage scenarios, including edge intelligence, metaverse, IoT networks, smart city, smart ocean, satellite, machine system, SAGIN, vehicular networks, and Wi-Fi sensing \cite{chen2024integrated}. As for edge intelligence, ISCC enables simultaneous high-quality environmental data sensing, delivery, and computation to maximize the inference/training accuracy under constraints on low latency and on-device resources \cite{wen2024integrated}. The task-oriented ISCC designs are further tailored to account for different AI applications  \cite{xing2023task}. As an AI-generated virtual world, the metaverse requires tremendous sensing, communication, and computation resources, where the ISCC can be applied to cope with the conflict between limited resources and user demands \cite{wang2024integration}. In IoT networks, the application of ISCC can improve the spectrum efficiency and reduce latency by enabling simultaneous sensing, communication, and computation \cite{li2023over}. In smart city, ISCC is expected to promote the development of information infrastructure \cite{liu2022communication}. In smart ocean, ISCC is applied for connecting marine objects in surface and underwater environments \cite{dai2022survey}. As for satellites, ISCC is proposed to cope with the complex and dynamic satellite environments and limited network resources \cite{zuo2024integrating}. In machine systems, the application of ISCC can realize low latency and high reliability of communication, highly accurate sensing, and fast environment adaption \cite{feng2021joint}. In SAGIN, ISCC is applied to meet the quality of service requirements of various applications \cite{zhang2024joint}. As for vehicular networks, ISCC is applied to increasing the sensing accuracy of the vehicular-equipped sensors while reducing the data communication and computation latency \cite{SUN2024Structural,LI2024cooperative,ZENG2024polarisation}. ISCC is also expected to be deployed in the scenario of Wi-Fi sensing for improving the data processing efficiency \cite{DENG2024statistical}.

\subsection{Challenges}\label{Sect:ChallengesISCC}
Although the ISCC design can achieve a higher performance ceiling, it encounters new challenges, as elaborated in the following.
\begin{itemize}
    \item \emph{Multi-objective Optimization}: In the case of multiple-task coexistence, different tasks have different types of design targets and metrics. However, it is challenging to design a way that can jointly optimize these design metrics since they have different units. For example, the communication throughput is measured by bits, AirComp performance is generally measured by MSE between the computation result and ground truth, and sensing is measured by CRB. Existing designs generally combine these design metrics in a weighted sum form with a manually pre-determined weight for each metric. This fails to accurately characterize the importance levels of different tasks, leading to a sub-optimal performance. It remains as a challenge to design an efficient and effective multi-objective optimization mechanism.
    \item \emph{Inference Management}: In the case of the coexistence of multiple waveforms (e.g., sensing, communication, and AirComp signals), interference should be carefully managed to guarantee the quality of each module.
    \item \emph{High Signal Design Complexity}: In the case of reusing one waveform to perform dual functional of sensing and communication (see, e.g., \cite{li2023integrated}) or triple functional of sensing, communication, and computation  (see, e.g., \cite{li2022inte}), low-complexity signal processing techniques are demanded.
    \item \emph{Complicated Resource Management}: The radio resource management (RRM) from a systematic view to jointly coordinate the sensing, communication, and computation modules is much more complicated, due to the tight coupling among them and the enlarged problem size \cite{wen2022task}.
    \item \emph{Lack of Unified Design Criteria}: Different from the existing communication systems with a common design goal (e.g., channel capacity and SNR), there is a lack of a unified design criterion for intelligent tasks with customized goals. For example,  federated learning tasks aim to achieve several goals of high testing accuracy, minimum convergence latency, data privacy, least energy consumption on devices, etc. (see, e.g., \cite{liu2022toward, wen2022federated}). This task-oriented property leads to a new challenge of finding suitable mathematical expressions to model the ISCC problem.
    \item \emph{Generalization}: Existing ISCC schemes are designed for specific tasks (e.g. federated learning, edge AI inference), for certain networks (vehicular or sensor networks), or under particular scenarios (coexistence of two or more specific tasks). There is a lack of an ISCC strategy that can be commonly utilized or easily transferred to other tasks, networks, or scenarios.
\end{itemize}
Many efforts have been put on ISCC to address these challenges. In this paper, we categorize these techniques into two types, i.e., signal design and network resource management, which are presented in Sections 
\ref{Sect:SignalDesign} and \ref{Sect:RRM} respectively.

\section{Signal Design for ISCC}\label{Sect:SignalDesign}

\subsection{Overview}


To support ISCC, the signal designs have been investigated including the beamforming and waveform designs. Specifically, the waveforms need to be designed to balance the radar sensing performance evaluated by CRB, the communication performance evaluated by SNR, and the computation performance evaluated by AirComp MSE between the aggregated data and the ground truth. The beamforming designs are further required to equalize the channels among mobile devices and suppress noise to improve the ISCC performance.

In the physical layer, ISCC can be categorized into three types as shown in Fig.~\ref{FigPhysical}. The first type designs independent signals dedicated for sensing, communication, and computation functionalities, which is known as the \emph{single-functional signal design}. However, the dedicated signals cause severe interference among different functionalities. To reduce the interference, the second type designs one signal for sensing and another for AirComp, which is defined as the \emph{dual-functional signal design}. As a further step, the third type namely the \emph{triple-functional signal design} realizes sensing, communication, and computation in one signal, which trades increased design complexity for interference cancellation. The technical details are elaborated as follows.

\begin{figure*}[ht]
\centering
\includegraphics[scale=0.45]{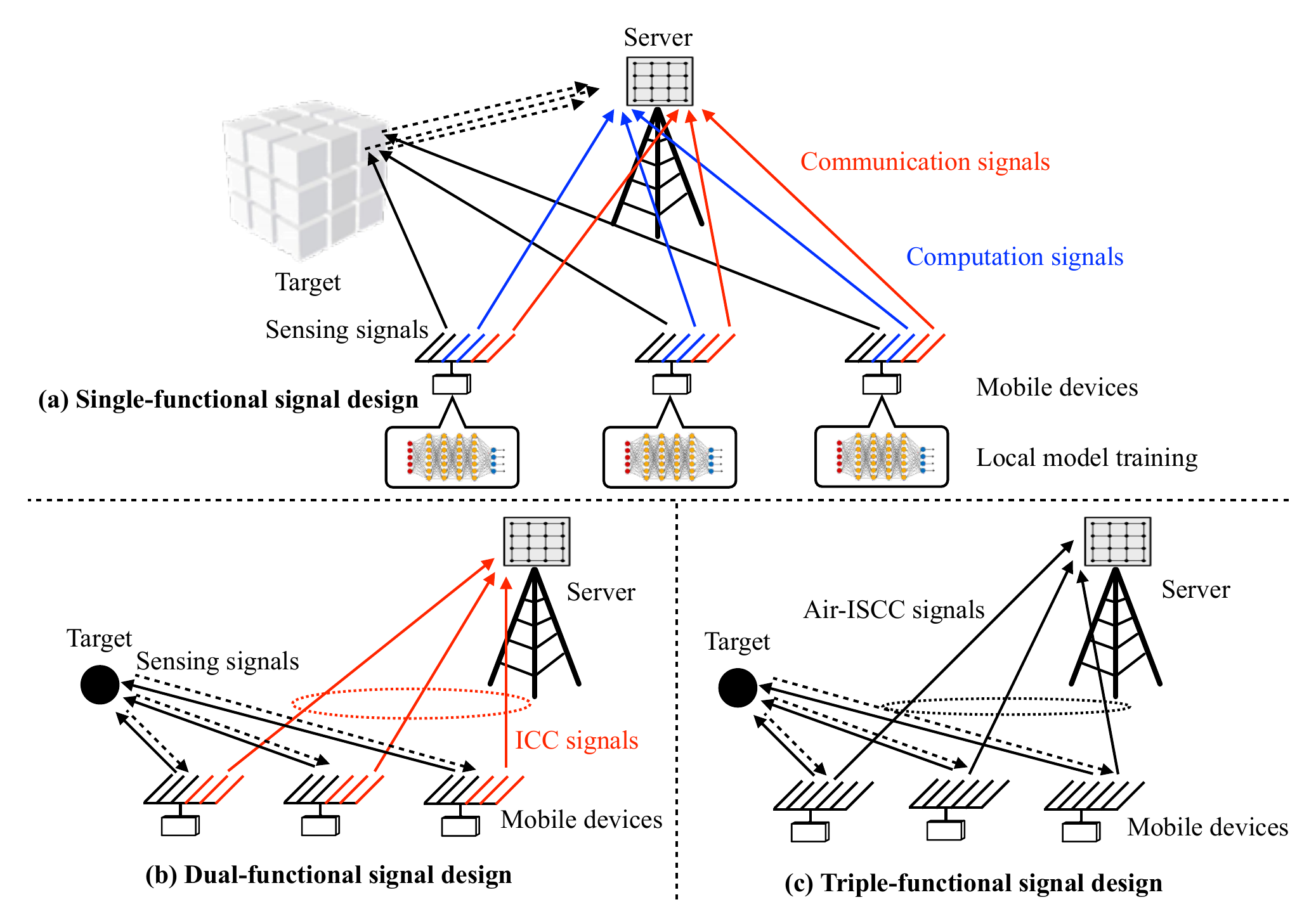}
\caption{ISCC signal designs.}
\label{FigPhysical}
\end{figure*}


\subsection{Multi-Functional Signal Design}

\subsubsection{Single-Functional Signal Design}
In \cite{qi2020integrated}, the independent communication signals and computation signals are transmitted simultaneously from multiple multi-antenna mobile devices (MDs), while the data is assumed to be sensed by the MDs. The communication signals are received by the server to recover the data with the existence of interference and channel noise, where the performance is evaluated by the signal-to-interference-plus-noise ratio (SINR). As for computation, The signals are superposed during the transmission and aggregated at the server, where the performance is evaluated by the MSE between the aggregated data and the ground truth. To improve the communication throughput and reduce the computation error, the transmit and receive beamforming designs for communication and computation are jointly designed. However, the neglect of the sensing process makes it less possible to be implemented in practice.

By taking the sensing process into consideration, the dedicated signals for sensing, communication, and computation are transmitted simultaneously from multiple single-antenna MDs in \cite{qi2022integrating}. The sensing signals are reflected by the targets and then received by the server, based on which the information of targets is estimated. The performance of sensing is evaluated by the MSE between the estimated target information and the ground truth. To reduce the interference between different functionalities, dedicated beamforming designs for sensing, communication, and computation have been investigated in \cite{qi2022integrating}. The algorithm complexity of the beamforming design is $\sqrt{2K}(n_1 K + n_1^3)$, where $n_1 = \mathcal{O}(K^2 M^2)$ with $K$ denoting the number of MDs and $M$ denoting the number of antennas at each MD. The computation signals carrying local learning results at each sensor are aggregated at the server to update the global learning model, where the performance is evaluated by the MSE between the aggregated local model parameters and the ground truth. To reduce the interference between different functionalities, dedicated beamforming designs for sensing, communication, and computation have been investigated. However, the interference among different signals still exists. To reduce the interference, a series of works explore the possibility of integrating the functionalities of sensing, communication, and computation in one common signal. 

\subsubsection{Dual-Functional Signal Design}
In the dual-functional signal design, the antenna array on each MD is divided to produce two distinct signals, one for sensing purposes and the other for AirComp \cite{li2023integrated}. The information of the sensing target is extracted from the echo signal received at the MD. At the same time, the gathered data is computed through the analog addition of signal waves during the data transmission from the MDs to the server. The MSE of the estimated target response matrix is used for evaluating the sensing performance, while the MSE of the computation results is used for evaluating the AirComp performance. Compared with the single-functional design where the interference between the signals for sensing, communication and computation needs to be managed, the dual-functional design only needs to manage the interference between the signals for sensing and AirComp. To improve the AirComp accuracy while guaranteeing the sensing performance, the sensing transmit beamformers, data transmit beamformers, and data aggregation beamformers are jointly designed. The impact of AirComp signal interference on radar sensing can be mitigated using statistical methods, while the interference of sensing signals on AirComp will result in deteriorated performance, which needs to be further dealt with. 

\subsubsection{Triple-Functional Signal Design}
As a further step of integration, ISAC and AirComp are combined in a triple-functional signal design, which is known as the \emph{integrated sensing, communication, and computation over-the-air} (Air-ISCC) \cite{li2022inte}. At each MD, the entire antenna array produces a single signal to realize sensing, communication, and computation functions. 
The signal sent by each MD serves as both the radar probing pulse and the data carrier, necessitating the design of a transmission beamformer on each MD. The target of radar sensing can be determined from the statistical data of the reflected signal using \emph{maximum likelihood estimation} (MLE). As for AirComp, the data aggregation beamformer is deployed at the server for equalizing the received signals. Compared with the single and dual-functional signal designs, the triple-functional signal design results in a more complex optimization problem for minimizing the AirComp error while guaranteeing the sensing performance. The desired design is achieved using a solving approach that employs semi-definite relaxation.

To improve the performances of radar sensing and AirComp, a series of beampattern designs are further investigated in \cite{wang2023inte}. Both the omnidirectional and directional radar beampatterns are designed, where the AirComp accuracy of the directional beampattern design is not as good as that of the omnidirectional one. 
Additionally, the balance between the optimal radar sensing beampattern and AirComp accuracy is explored. Accounting for the total power and per-antenna power budget constraints respectively, the corresponding beampattern designs for balancing the trade-off are proposed in \cite{wang2023integrated}.

As a promising technique for improving spectrum efficiency, Air-ISCC can be applied in a series of scenarios including target localization, vehicular networks, and federated learning \cite{li2023over}. In federated learning, the Air-ISCC signals are used to simultaneously sense the information of the target for local model training and deliver the local training results to the server via AirComp for global model update \cite{xing2023task}. Moreover, an IRS is further deployed to assist the AirComp \cite{zheng2023federated}.

\subsection{Discussion}
The ISCC signals can be applied in a series of scenarios including target location estimation, vehicular networks, and federated learning \cite{li2023over}. As for target location estimation, the location of the target is estimated by multiple MDs based on the information of relative distance and angle estimated from the reflected radar signals, as well as their own locations. Meanwhile, each MD delivers its previous locally estimated location of the target to the server simultaneously, and thus the server obtains the averaged estimated target location via AirComp. In vehicular networks, the ISCC signals are expected to train the beam for vehicular tracking and predicting. In federated learning, the ISCC signals are used to simultaneously sense the information of the target for local model training and deliver the local training results to the server via AirComp for global model update \cite{xing2023task}. The single-functional design has the advantage of simple waveform design for dedicated functionality, but suffers from severe interference. The dual-functional design mitigates the interference between communication and computation interference by exploiting the AirComp property, while the interference between sensing and AirComp still exists. The triple-functional design gets rid of the interference among different functionalities, but at sacrifice of complex waveform design.

\section{Network Resource Management for ISCC}\label{Sect:RRM}

\subsection{Overview}
Apart from signal designs for ISCC, another main research stream of ISCC aims to enhance the network resource utilization via resources (e.g., time, bandwidth, and energy) sharing. Particularly, network resource utilization refers to the cost of network resources to achieve a given task performance. As shown in Fig. \ref{RRM_ISCC}, there are two ISCC resource allocation paradigms. One paradigm is joint RRM for task coexistence, which allows the coexistence of sensing, communication, and computation tasks, leading to loose coupling among the three modules as shown in Fig. \ref{RRM_ISCC}(a). In this paradigm, each task has its own goal. Dynamic network resource allocation among different tasks can be conducted to achieve multiple goals (e.g., communication capacity, computation accuracy, and sensing SNR) simultaneously. The other paradigm is task-oriented ISCC resource allocation for complicated tasks that involve sensing for information acquisition, computation for information processing, and communication for information transmission. 
In these tasks, the three modules are tightly coupled as shown in Fig. \ref{RRM_ISCC}(b). The reasons are as follows. On one hand, the overall task performance depends on the qualities of the three modules, e.g., the sensing and communication SNR, the computation latency, and the quantization error. However, they should compete for resources to enhance their own qualities. On the other hand, there is tight dependency among them in general. For example, the size of the obtained sensory data decides the computation load, while the size of on-device computational results further determines the communication load. In the sequel, the above two paradigms are elaborated. 
\begin{figure*}[t]
\centering
\includegraphics[width=1\textwidth]{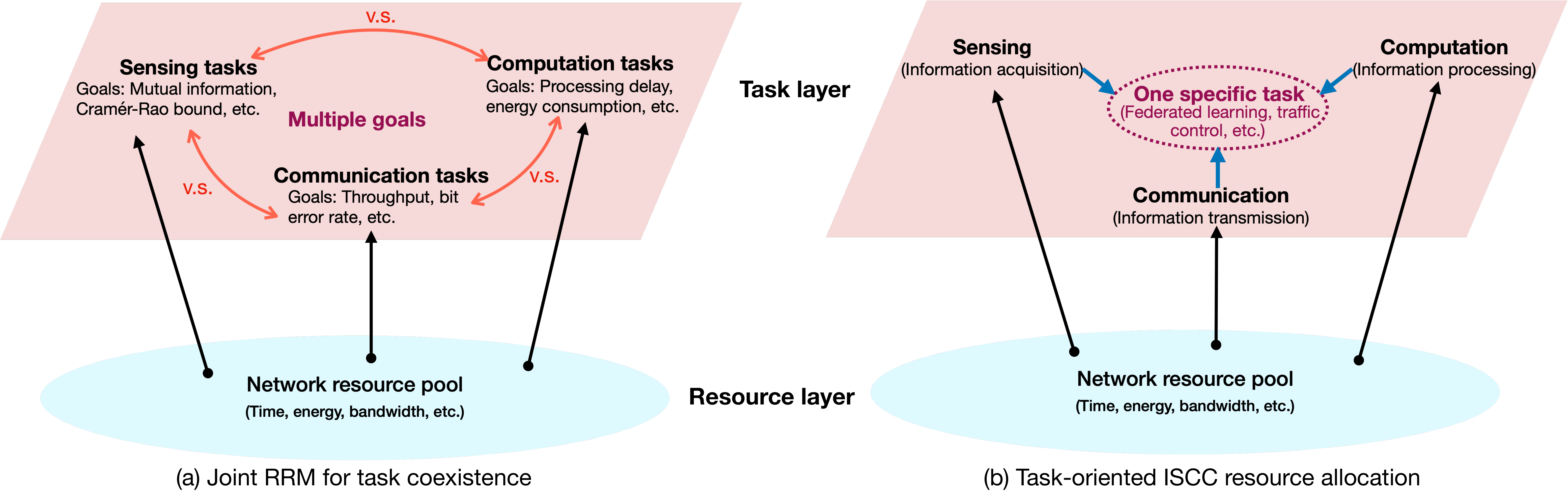}
\caption{Two ISCC resource management paradigms.}
\label{RRM_ISCC}
\end{figure*}

\subsection{Joint RRM for Task Coexistence}
In this paradigm, sensing, communication, and computation tasks coexist and network resources are jointly and adaptively allocated among them to achieve multiple goals at the same time. In general, sensing and communication tasks are executed in physical layers, while computation tasks are realized in upper layers by e.g., MEC. In this paradigm, each type of task adopts its traditional performance metric. As an example, in \cite{zhao2022radio}, the sensing performance is measured by the \emph{mutual information} (MI), the communication task aims to achieve a higher transmission rate, and the computation performance is evaluated by the processing delay. In the following, three typical example cases of this paradigm are discussed, i.e., the coexistence of sensing, communication, and computation tasks, the coexistence of sensing and MEC tasks, and the coexistence of ISC and MEC tasks.

\subsubsection{Coexistence of Sensing, Communication, and Computation Tasks}

In this case, sensing, communication, and computation tasks coexist and compete for network resources to achieve their own targets. Particularly, in \cite{zhao2022radio}, ISAC serves as a supporting technique to allow hardware and spectrum sharing between the communication and the sensing tasks. A Cobb-Douglas utility function that combines three metrics, i.e., sensing precision, communication capacity, and computation capability, is adopted and maximized by the joint design of device association and sub-channel assignment. Besides, the authors in \cite{cheng2022optimized} investigate a time-division multiplexing ISCC system, where FMCW is used as the sensing waveform and orthogonal time slots are assigned to the sensing, uplink/downlink communication, and computation processes. Targeting communication capacity maximization and computation offloading latency minimization while guaranteeing the sensing distortion is under a limited threshold, a joint resource allocation, time partitioning scheme, computation task processing mode selection, and target sensing location selection, are proposed. In \cite{xu2022irs}, the IRS is applied to assist the ISAC and computation task offloading, where both partial offloading and binary offloading are considered. A performance metric called weighted throughput capacity, which comprises the radar sensing capacity, communication throughput, and computation throughput, is proposed and maximized via joint beamforming design and time allocation. Furthermore, in \cite{feng2021joint}, the ISCC framework is designed for intelligent machine-type communication (IMTC) networks to achieve highly reliable communication with low latency, highly accurate sensing, and fast data processing. In IMTC, intelligent machines (IMs) are expected to execute complex data processing tasks in rapidly varying environments, which demands IMs to conduct real-time environmental sensing and cooperation with each other \cite{mahmood2020six}. To this end, a closed-loop is formulated where the sensed data is exploited as a priori information to improve the communication performance, the communication techniques are designed to support the intelligent computation, and the computation results in turn facilitate the ISAC process. Moreover, the resource allocation among sensing, communication, and computation tasks in vehicular networks is investigated in \cite{yang2024deep} under the principle of maximizing the achievable data rate while guaranteeing sensing and computation performance.

\subsubsection{Coexistence of Sensing and MEC Tasks}

In this case, the sensing and the MEC tasks compete for radio resources to achieve their respective goals. ISAC technique is used to enable hardware and radio resource sharing between sensing and communication.  In \cite{10045764}, a UAV-supported ISCC framework is proposed, where a treble-functional UAV senses a target by using the unified beam waves, processes sensory data locally, and offloads its computational tasks to ground APs. By characterizing the Pareto boundary between the sensing beampattern gain and the computation capacity, a fundamental trade-off is revealed. Specifically, by applying the semidefinite programming method and the concave-convex procedure technique, a joint optimization algorithm is proposed and depicts the performance region, as well as enhancing the ability of sensing, computing, and communication of UAVs. Besides, an architecture that integrates communication, radar sensing, and MEC, is developed in \cite{9729765}, where user terminals conduct computation offloading and radar sensing over the same radio resource blocks by using DFRC techniques and MIMO arrays. To jointly optimize the performance of MIMO radar beampattern design, computation offloading energy consumption, the individual transmit precoding for radar, and resource allocation of communication and computation, a multi-objective optimization problem is formulated. 
The authors in \cite{9996408} develop a non-orthogonal multiple access (NOMA)-aided joint communication, sensing, and multi-tier computing (JCSMC) framework, where the BS can not only sense target areas and offer MEC services to the covered mobile devices simultaneously but also offload some computation tasks to cloud servers to improve the computation efficiency. The potential benefits of utilizing NOMA in this framework are investigated. Both the transmit beamformer and computing resource allocation are jointly optimized to maximize the computation rate in two computation offloading modes, i.e., partial and binary offloading modes. An IRS-assisted ISCC system is proposed in \cite{wan2024reconfigurable} to support multiple devices simultaneously sensing for target detection and computation offloading. The sensing quality and the communication throughput for computation offloading are enhanced by the IRS. Moreover, a multi-point aided ISCC framework is proposed in \cite{dong2024coordinated}, where multiple base stations cooperatively conduct target detection via sensing and serve computation offloading for multiple devices simultaneously. 



\subsubsection{Coexistence of ISC and MEC Tasks}

In this type, an ISC task competes with an MEC task for network resources to enhance respective computational performance. Particularly, in \cite{he2022integrated}, computation resource competition between an AI model inference task and an MEC offloading task in an edge server is investigated where sensing serves as the supporting technique for completing the AI model inference task and communication serves as the supporting technique for the MEC task. A novel action detection module is first proposed for alleviating the network resource competition by detecting the status of the sensing target. If the sensing target is static, no further computation is required for the AI model inference task and hence reduces the communication overhead at the edge server.  Then, the inference accuracy is maximized while guaranteeing the quality-of-service (QoS) requirements of the MEC task via designing an optimal strategy of joint resource allocation and static status threshold selection.  

As well as the above three cases, authors in \cite{deng2024integrated} propose a more complex framework where an edge server simultaneously provides the ISCC-supported DNN splitting inference services and the communication services for multiple UAVs. The average sum rate of all UAVs is maximized under the latency and accuracy guarantees of the DNN splitting inference tasks. 




\subsection{Task-oriented ISCC Resource Allocation}
A massive number of intelligent devices are deployed at the network edge to complete various kinds of tasks for enabling intelligent connection of everything \cite{zhang2022towards}. In this paradigm, network resources are allocated among the sensing, communication, and computation modules of a complex task to achieve its customized goal. The fulfillment of complex tasks requires sensing for information acquisition, communication for information transmission, and computation for information processing, such as auto-driving tasks, federated learning tasks, and edge inference tasks \cite{wen2024integrated,shi2023task}. In other words, it is desirable to generate and deliver the most valuable information that can be used to effectively and efficiently complete an intelligent computation task \cite{zhang2022towards}. However, the qualities of the three modules decide the performance of these tasks but the three modules compete for the limited radio resources. Moreover, there is a lack of a uniform design objective for these tasks. For example, instead of targeting at throughput maximization in conventional communication tasks, a federated learning task concerns improving learning accuracy while reducing training latency. To this end, various task-oriented ISCC resource allocation schemes are developed, as discussed below.

\subsubsection{ISCC for Federated Learning Tasks}
In each training iteration of an online federated learning task, an edge server first broadcasts a global model to all participating devices. Then each device acquires training data samples via sensing, calculates the local gradient via computation, and uploads the local gradient to the server via communication \cite{liu2022toward,wen2024AirCompISCC,liang2024federated}. The server finally aggregates all local gradients to update the global model. In this procedure, on-device sensing, computation, and communication are tightly coupled for the following reasons. The learning performance, e.g., convergence rate and per-iteration loss function degradation, depends on the per-iteration gradient distortion, which is decided by the number of obtained and processed data samples, the sensory data distortion (sensing SNR), and the transmission distortion (communication SNR). However, it is hard to enhance all these metrics due to the limited network resources \cite{wen2024AirCompISCC}. To this end, the learning performance guided ISCC resource allocation schemes are desirable. In \cite{liu2022toward}, by operating sensing, on-device computation, and communication in a time-division manner and using FMCW for sensing, an ISCC-based federated learning framework is proposed with the goal of accelerating the convergence rate. However, the influence of sensing noise on the convergence performance is ignored in  \cite{liu2022toward}, which limits the application of its proposed scheme especially in the case of hostile sensing channels and limited network resources. This scheme is further extended to the case of UAV-assisted federated learning \cite{10446598}. To characterize the influence of sensing noise, the authors in \cite{wen2024AirCompISCC} theoretically characterize the influence of sensory data distortion, communication SNR, and the number of sensed and processed data samples on the convergence rate. Guided by this analysis, an ISCC resource allocation scheme is proposed for over-the-air federated learning.


\subsubsection{ISCC for Edge AI Inference Tasks}
A real-time inference task involves sensing for data acquisition, communication for data sharing, and computation for data processing and intelligent decision-making \cite{wen2024integrated}. The inference accuracy depends on the distortion caused during the three processes, e.g., sensing clutter and noise, quantization distortion, and channel noise, but the three processes compete for network resources to enhance their own qualities. Therefore, to enhance the inference accuracy with critical latency requirements, several ISCC resource management schemes are proposed. In \cite{wen2022taskaircomp}, the AI model is split into two parts, the front-end part deployed at devices is used for feature extraction from sensory data. The backend part with intensive computation deployed at an edge server completes the inference task by receiving all local feature vectors from all devices. Specifically, multiple devices sense the same wide view of the source target and obtain a local distortion-corrupted version of the ground-true sensory data. Then, each device extracts a low-dimensional local feature vector from its local sensory data using a principal component analysis based linear extractor. An edge server aggregates all local feature vectors via AirComp to obtain a denoised global feature vector, which is input into the remaining AI model for finishing the downstream inference task. This work directly adopts the inference accuracy as the design objective rather than the conventional metric of MMSE, as the latter ignores the feature elements with the same distortion but heterogeneous contributions to the inference accuracy. Since the instantaneous inference accuracy lacks a mathematical model, a surrogate metric called discriminant gain is applied for classification tasks \cite{9955582}. It measures the distances between different classes in the feature space based on the symmetric KL divergence. Particularly, with a higher value of discriminant gain, different classes are better separated in the feature space, leading to a higher achievable inference accuracy. Besides, the communication efficiency is enhanced using AirComp. The above work is extended to the case with sensing noise suppression \cite{zhuang2024integrated}, the decentralized case \cite{zhuang2023decentralized}, and the case of unbalanced classification accuracy \cite{10586815}. To further deal with the scenario where different devices sense disjoint narrow views of the source target, a task-oriented sensing, feature quantization, and feature transmission technique is proposed in \cite{wen2022task}, where the local feature vectors of different devices are cascaded at the server to form a global one. Based on \cite{wen2022task}, an ISCC-based selection approach is proposed in \cite{10631278} to choose the best from the modes of on-device, on-server, and edge-device cooperative. Moreover, ISCC-based joint device scheduling and resource allocation methods are proposed for enhancing the inference accuracy while preserving the data privacy \cite{wang2023device} and simultaneously completing multiple inference tasks \cite{10669354}.

\subsubsection{ISCC for Vehicular Networks}
In this case, task-oriented ISCC resource allocation schemes are proposed to enhance the system performance, e.g., achieve low latency for auto-driving \cite{zhang2021joint} and address road traffic congestion \cite{9665372}. Specifically, in \cite{zhang2021joint}, the message and resource allocation are investigated for cooperative perception in fog-based vehicular networks. The satisfaction of cooperative perception (SCP) is firstly characterized based on the spatial-temporal value and latency performance. To maximize the sum SCP, the sensing block message, communication resource block, and computation resource are jointly allocated under the maximum latency and sojourn time constraints of vehicles. Besides, in \cite{9665372}, the ISCC technique is regarded as an appealing control tool for road traffic congestion. The evolution of road traffic congestion control is divided into two sequential stages, i.e., the sensing and the communication stages. In the former stage, congestion control strategies experience rapid growth due to the accessibility of traffic data. In the latter stage, with advances in vehicular networks, big data and AI, the improvement of the ability to collect finer data is significant. This paper also introduces how existing congestion detection techniques evolve with the continuous development of sensing, communication, and computation capability in the networks. Moreover, the sensing, communication, and computation are modeled in the scenario of continuous and random task arrival in vehicular networks and a collaborative sensory data fusion framework is proposed in \cite{zhao2024IoV} for maximizing the task completion rate.  


\subsubsection{ISCC for Mobile Crowdsensing}
In the paradigm of mobile crowdsensing, devices sense the environment to collect data via the equipped sensors and offload the data to a server after pre-processing to reduce the communication load for further processing. The design goal is to maximize a data utility function, which depends on the acquired data quantity and quality \cite{9927216}. Due to hostile wireless channels, limited spectrum, energy, and computation resources, the transmission or processing of sensing data may be unsuccessful in wireless edge networks. To this end, an ISCC framework for multi-dimensional resource-constrained crowdsensing systems is proposed in \cite{9927216} to optimize the overall performance by jointly selecting users, allocating bandwidth, and designing the strategies. It should be emphasized that the data utility function in this paradigm cannot well characterize the customized goals of the federated learning and edge AI inference tasks, although the sensing, computation, and communication procedures are similar. For example, the data utility function fails to quantify the distortions caused during sensing, computation, and communication on the learning or inference performance as well as the long-term influence on the convergence rate.

\subsubsection{ISCC for Sensor Networks}
In \cite{1285548}, a semantic model and a hybrid perceptive frame approach are proposed for the control of ISCC-based reconfigurable mobile sensor networks. First, the mobile sensor network is modeled as an ISCC-based dynamic system. Then, the relation between different sensors and the corresponding multi-modal sensory information is characterized by a proposed semantic model. On this basis, by describing the hybrid nature of cooperative sensor networks and the integration of sensing, communication, and cooperation as perceptive references and semantic representations, the mobile sensor network can autonomously reconfigure itself adapting to the varying operating conditions.

\subsubsection{ISCC for Robotics}
An ISCC-based unmanned UAV-robot system is proposed in \cite{fang2024sensing}, where the UAVs enabled with the functions of sensing, communication, and computation, are deployed for controlling robotics for task execution. Specifically, UAVs process the collected real-time sensory data to generate control commands, which are further transmitted to robotics for execution.  Such workflows require the cooperation and coordination of sensing, computation, communication, and control (SC$^3$). To this end, a task-oriented closed-loop SC$^3$ optimization approach is proposed to simultaneously overcome the scarce channel bandwidth, limited energy supply, and sensing distortion for minimizing total linear quadratic regulator cost.



\subsection{Discussion}
By enabling resource sharing, coordination, and joint optimization among sensing, communication, and computation under multiple objectives or a customized goal, network resource utilization and task performance can be enhanced. However, the problems of RRM in ISCC systems are more complex than those of the separated designs, as more tightly coupled variables should be jointly optimized. Moreover, a customized scheme should be designed for different scenarios and tasks. The current ISCC RRM strategies have relatively low generalization capability, leading to high implementation costs in different settings. To this end, more general, simple and efficient ISCC RRM strategies are demanded in the future. On the other hand, ISCC RRM is suitable for critical tasks and cases with extreme performance requirements or limited network resources. Otherwise, the separated RRM designs are sufficient.

As well as the tasks mentioned previously, there are many other tasks that involve the coordination of sensing, communication, and computation to enhance performance with the limited network resources, e.g., holographic communication, auto-driving, intelligent agriculture, and smart healthcare.  In these tasks, the sensing, communication, and computation processes are tightly coupled via competing network resources but cooperate to achieve a customized goal. However, the design of task-oriented ISCC RRM for these tasks demands further exploration.

\section{ISCC in the Future}

In this section, three future implementation examples of ISCC are first provided, including the potential ISCC techniques for deploying digital-twins platforms at the network edge and the potential ISCC applications in two future advanced networks, i.e., the computing power networks and the SAGINs. Then, several critical unresolved issues that are significant to efficiently implement ISCC in the future are discussed.

\subsection{ISCC for Digital-Twins Enabled Wireless Networks}

\begin{figure*}[h]
\centering
\includegraphics[width=0.75\textwidth]{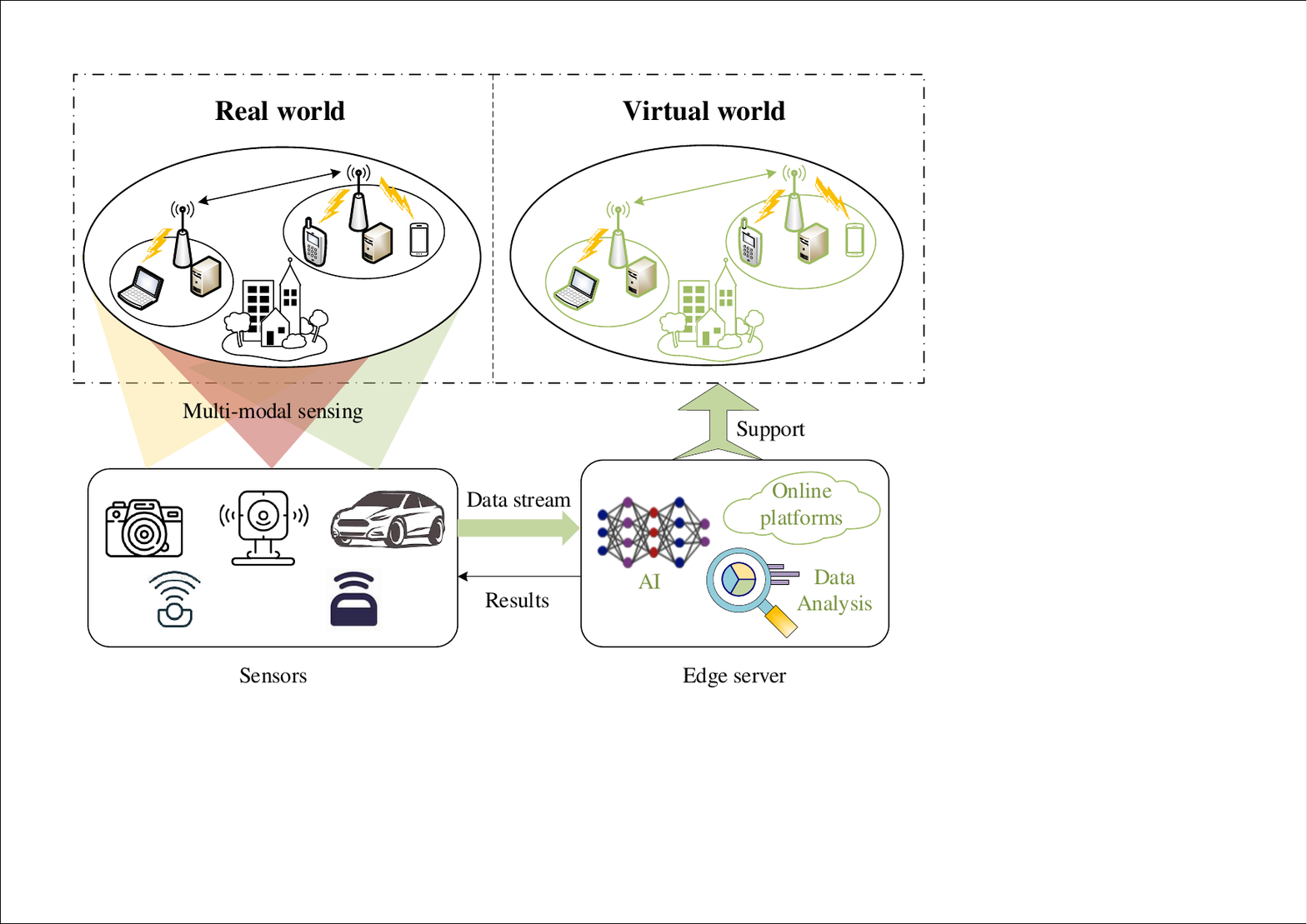}
\caption{An ISCC-based digital-twins network.}
\label{DT}
\end{figure*}
A digital-twins system refers to constructing virtual or digital representations of the physical objects and systems via the techniques of high-quality distributed sensing, precise 3D map construction, real-time tracing, deep machine learning, and so on \cite{alkhateeb2023real}. It targets generating simulation data that cannot be distinguished from its real-world counterparts, conducts data analysis, and makes decisions based on the generated data \cite{make5030054}. Digital twins have been widely used in many areas e.g., intelligent manufacturing, climate change, healthcare, and economic crises \cite{make5030054}. Particularly, since digital twins have a powerful ability to simulate complex wireless environments, it has attracted more and more attention to deploy digital twins at the network edge for addressing many issues including network complexity control \cite{wu2021digital}, network performance enhancement \cite{khajavi2019digital}, disruptions prediction and prevention \cite{bhandal2022application}, network topology visualization \cite{hui2022digital}, and supporting MEC and industrial IoT networks \cite{alkhateeb2023real}. 

However, establishing a digital-twins network is challenging. It desires massive sensory data of multiple modalities for acquiring ground-true information of real wireless networks, high communication throughput for transmitting sensory data or the corresponding extracted information, and powerful computation capability for data generation, analytics, and decision-making. To tackle these challenges, an ISCC-based digital-twin network is proposed as shown in Fig. \ref{DT}, where many sensors of different types are deployed for perceiving a real wireless network to obtain ground-true information, and the digital-twins system is deployed at an edge server for executing all complex and computation-intensive algorithms to construct the virtual representation. Based on such a digital-twins network, two ISCC techniques are discussed. One is a hybrid learning platform for training AI models that are used for constructing digital-twins systems. The other is a real-time ISCC technique to leverage the established digital twins for making accurate and fast decisions.

\begin{figure*}[h]
\centering
\includegraphics[width=0.78\textwidth]{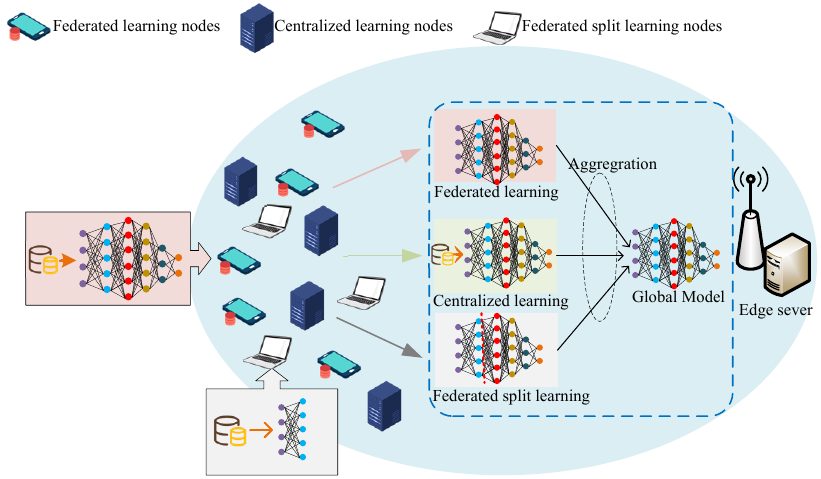}
\caption{An ISCC-based hybrid AI model training platform for establishing digital twins.}
\label{HybridPlatForm}
\end{figure*}

\subsubsection{Hybrid AI Training Based Digital-Twins Network Construction}
One significant step for establishing the digital-twins network in Fig.\ref{DT} is training many AI models for accomplishing many intelligent tasks, by using a large amount of multi-modal sensory data from many distributed sensors. However, different sensors have heterogeneous resources like energy and different capabilities in terms of sensing, communication, and computation. Besides, different sensors have different data privacy preservation requirements. To deal with this heterogeneous digital-twins network, a hybrid AI training platform that integrates federated learning, federated split learning, and centralized learning, is proposed to meet various requirements of different sensors, as shown in Fig. \ref{HybridPlatForm}. For example, for the sensors with critical privacy concerns and high computation and communication capabilities, federated learning is adopted. For the sensors with high privacy preservation requirements and low computation and communication abilities, federated split learning is performed. For the sensors without privacy concerns and with limited communication and computation resources, they can directly send the raw sensory data to the edge server for centralized learning. 

In each training iteration, the three training schemes are conducted in parallel. In federated learning, the global model is shared to the corresponding sensors for local training based on the obtained online and offline local data samples. In federated split learning, the global model is partitioned into a front-end part and a back-end part. The front-end part is shared to all involved devices for executing the forward pass using the online and offline sensory data and all local front-end parts are then transmitted back to the server for completing the downstream training. The centralized learning is conducted at the edge server by continuously collecting the real-time raw sensory data from the corresponding sensors. At the end of the training iteration, the updated models of the three training schemes are further aggregated to update the global model. 

The design object of such hybrid AI model training is to minimize the long-term resource consumption (e.g., energy consumption) of all devices to extend their lifetimes under the constraints of high learning accuracy, low training latency till convergence, and the various privacy requirements of different sensors. For simultaneously training multiple AI models in the hybrid platform, resources (e.g., time, energy, bandwidth) allocation needs to be performed among different tasks via joint design of sensing, communication, and computation. 

\subsubsection{ISCC Based Digital-Twins Network Application}
One key function of deploying digital twins at the network edge is making accurate and fast decisions for many network tasks such as resource allocation, multiple access, and computation offloading. To achieve this goal, it's desirable to collect the real-time sensory data from the sensors. Furthermore, the obtained sensory data should be sufficient and important to accomplish the specific task, as well as having high quality which means it cannot be polluted during the sensing, transmission, and processing. To this end, ISCC techniques should be designed, that should jointly consider many factors. One is the sensor's sensing capability including sensing range, data quality, sensing form (e.g., in an active way like radar sensors or in a passive way like cameras), the importance level of the obtained data to the task, etc. Other factors include the computation and communication abilities, the channel gains, battery {lives}, maximum transmit power, and the permitted latency for finishing the task. 

For simultaneously completing multiple decision-making tasks, the ISCC design is more complicated since different tasks compete for network ISCC resources. For saving network resources, an opportunistic sensor scheduling scheme can be designed to select the sensors, that can obtain real-time and high-quality data that is important to all tasks.

\subsection{Computing Power Networks Supported ISCC}
Benefiting from the rapid advancement of cloud computing, MEC, and smart devices, ubiquitous computation resources are deployed in the networks. Traditional network architectures are designed from the perspective of high communication efficiency and cause computing power island effect.  To overcome this shortage and to better leverage these distributed computation resources, a new computing network paradigm, called computing power networks, is proposed \cite{tang2021computing,yukun2024computing}.  It aims at coordinating the ubiquitous computing nodes in networks for providing more flexible computing power scheduling. For adapting to the heterogenous computing capabilities of different clouds, edge servers, and mobile devices, the micro-service architecture utilizes fine-granularity computing blocks to build up different complex tasks, where the computing blocks can be assigned to different devices according to their computing powers \cite{mendoncca2019developing,jang2021microservice,zeb2023toward}. For example, a micro-service based video analytics framework is proposed in \cite{yang2022latency}  for saving the computation overhead, where different applications like criminal tracking, traffic monitoring, and urban management, can share universal basic computing blocks for processing a same video, including frame decode, object detection, etc. Based on the micro-service architecture, two potential ISCC applications are investigated in computing power networks as follows. 

\begin{figure*}[h]
\centering
\includegraphics[width=1\textwidth]{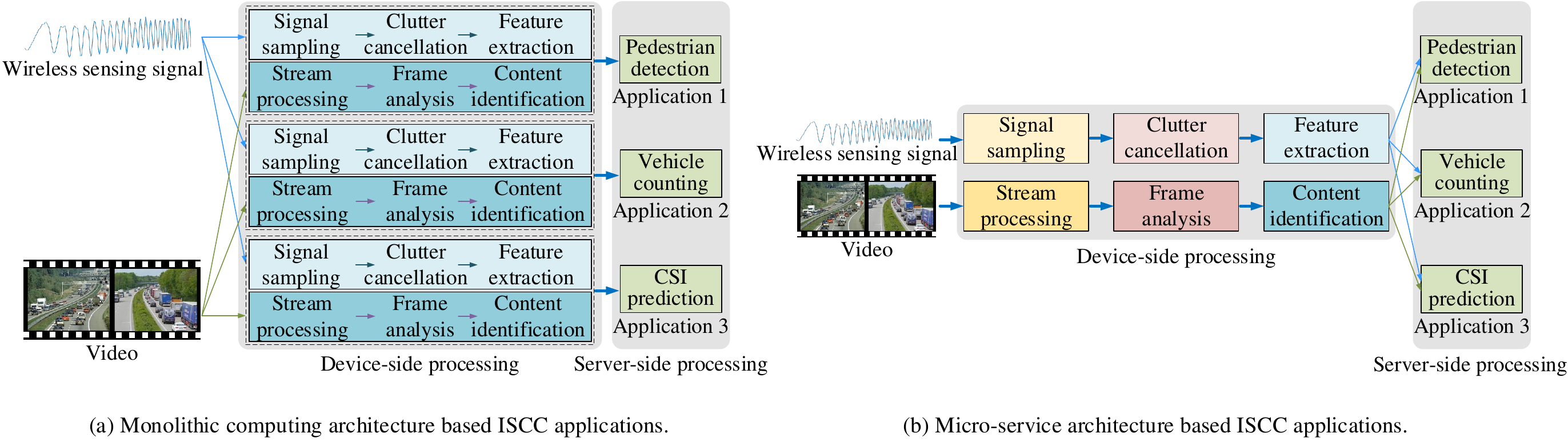}
\caption{Comparison of ISCC applications in monolithic computing and micro-service architectures}
\label{Fig:ISCC_CPN_1}
\end{figure*}
\subsubsection{Computing Sharing for Various ISCC Tasks}

As shown in Fig. \ref{Fig:ISCC_CPN_1}, an edge-device cooperative framework is deployed for supporting several ISCC tasks, where sensing and some basic sensory data processing for feature extraction are conducted at device and the DNNs based complicated computation tasks are performed at edge server. As a result, this framework can enjoy three-fold benefits of privacy preservation via avoiding raw sensory data transmission, low communication due to low-dimensional feature extraction, and computation offloading. In Fig. \ref{Fig:ISCC_CPN_1}, two computing architectures are compared for completing multiple ISCC tasks. In the conventional monolithic computing architecture as shown in Fig. \ref{Fig:ISCC_CPN_1}(a), each task has its own individual sensory data processing streams. The intermediate computing results cannot be shared among different tasks. Consequently, it leads to a severe waste of computing resources as well as increasing storage and communication costs. To tackle these issues, the micro-service architecture divides the signal processing of each task into many fine-granularity computing blocks as shown in Fig. \ref{Fig:ISCC_CPN_1}(b). Different tasks can share the same computing blocks, e.g., the blocks of signal sampling, clutter cancellation, and feature extraction for wireless sensing signal and the blocks of stream processing, frame analysis and content identification for video. This significantly saves the computing and storage overhead as well as reduces the communication load for transmitting the on-device processing results to the edge server.

\subsubsection{Directed-graph-based Joint Scheduling of Sensing, Communication, and Computation}

In computing power networks, the fine-granularity computing blocks for completing to make full utilization of the computing resources in various devices. Thereby, the completion of ISCC tasks requires cooperation of multiple nodes. For enhancing the performance of ISCC tasks while saving the resources in terms of sensing, computation, and communication, a directed graph model can be constructed. 
In the graph, there are two kinds of devices or called nodes. One is the sensing nodes with the capability of sensing, on-device computation, and communication. The other is the computing nodes with the ability of communication and computation. A target can be sensed by multiple sensing nodes and same computing blocks are deployed at different computing devices. As a result, the fulfillment of an ISCC task has multiple choices of paths. To be specific, a directed edge from node $a$ to node $b$ means that node $a$ transmits its computing results to node $b$ for downstream processing. In each path, sensing and all required computing blocks are sequentially performed for completing the task. Whether there is {a} directed edge between two nodes depends on whether a wireless communication link can be constructed between them for reliably transmitting the intermediate computing results based on the factors like transmit power, bandwidth, time, and channel gain. In order to enhance the task performance while saving the sensing, communication, and computation costs, it's desirable to find the best path in the directed graph.

\subsection{Space-Air-Ground Integrated Networks Supported ISCC}

By integrating satellite, aerial, and ground communication systems, SAGIN has been a promising solution for providing infrastructure-free environments with communication and intelligent services such as precision agriculture, environmental monitoring, and wildlife monitoring \cite{liu2018space,gong2024intelligent}. On one hand, compared with conventional terrestrial networks, SAGIN can provide global coverage especially for remote areas by using UAVs as relays and links between UAVs and satellites for providing backhaul connectivity. On the other hand, satellites are equipped with many remote sensors for detecting and monitoring the characteristics of earth systems from the space in a global perspective. This enables real-time and accurate {decision-making} for many global mission critical such as disaster assessment. Based on the satellites' two types of functions mentioned above, two SAGIN-supported ISCC frameworks are introduced below.

\begin{figure}[h]
\centering
\includegraphics[width=0.42\textwidth]{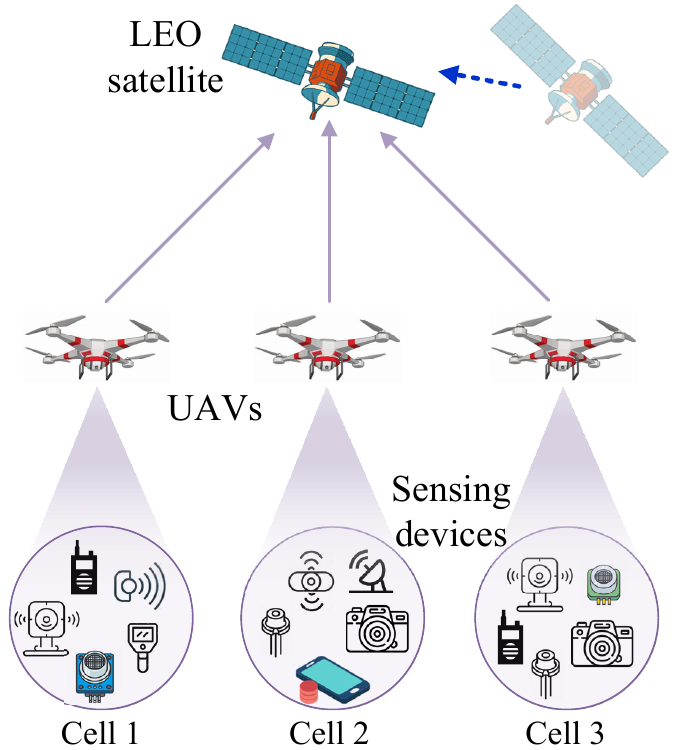}
\caption{Terrestrial sensing based ISCC framework in SAGINs.}
\label{Fig:Satellite_1}
\end{figure}
\subsubsection{Terrestrial Sensing based ISCC Framework in SAGINs}
Consider a hierarchical SAGIN as shown in Fig. \ref{Fig:Satellite_1}, where UAVs relay the sensory data from many terrestrial sensing devices to satellites for completing real-time intelligent tasks. Since computation migration between different satellites is difficult due to the high communication load and weak channel link gains between different satellites, only one satellite is selected for completing one task. Each UAV accesses several sensing devices within a disjoint coverage. The communication links between UAVs and the terrestrial sensing devices in their coverage are strong but the links between UAVs and satellites are weak due to the large path loss caused by the long distance. To tackle this issue, all UAVs collect sensory data from the sensing devices in their own coverages, pre-process the collected data for feature extraction and compression to save communication overhead, and forward the extracted and compressed feature. The satellite conducts a global fusion of the extracted and compressed features from all UAVs for executing the task. Compared with traditional terrestrial networks, SAGINs have the ability to connect the sensing devices in remote areas and can support much larger sensing ranges for more complicated ISCC applications. In this framework, ISCC designs are demanded to jointly coordinate the terrestrial sensing, communication from sensors to UAVs, signal pre-processing on UAVs, communication between UAVs to the satellite, and the on-satellite computation to achieve high task performance.

\begin{figure}[h]
\centering
\includegraphics[width=0.42\textwidth]{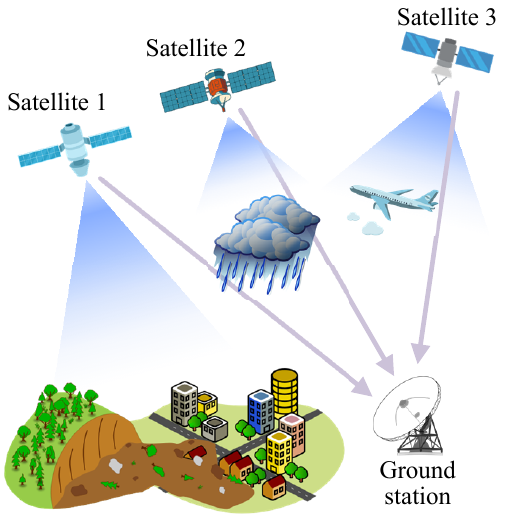}
\caption{Remote sensing based ISCC framework in SAGINs.}
\label{Fig:Satellite_2}
\end{figure}
\subsubsection{Remote Sensing based ISCC Framework in SAGINs}
Another main function of satellites is remote sensing by being equipped with different types of sensors. As shown in Fig. \ref{Fig:Satellite_2}, by obtaining different modalities of sensory data such as panchromatic images, multispectral images, thermal imaging infrared data, and radar sensory data, and transmitting the processed features to the ground station, many real-time tasks can be accomplished such as precision agriculture and climate change detection. To establish such systems, joint design of multi-modal sensing across multiple satellites, on-satellite computation, and satellite-ground communications are required. As well as addressing the same challenge of weak and fast varying channel links as the terrestrial-sensing-based ISCC framework, there are new design requirements including satellite scheduling for selecting the best sensing views and areas of the task, the multi-modal data alignment across different satellites, and the multi-modal data fusion schemes. For example, the data fusion schemes can be categorized into three types. The first is that satellites directly transmit the raw sensory data to the ground station. This scheme can keep the most sensing information but encounters severe communication bottlenecks for transmitting high-resolution sensory data. The second scheme is called early fusion, where each satellite transmits an extracted low-dimensional feature map to the ground station by saving communication load but sacrificing a part of the information. The last one is late fusion, where on-satellite decisions are made and only the results are fused at the ground station via e.g., a majority vote. In this scheme, only very little information is shared to the ground station but high communication efficiency can be achieved. Besides, since late data fusion shares minimum information to the ground station, it can be conducted when satellites and the ground station belong to multiple institutions.

\subsubsection{Challenges}
The implementation of SAGIN-supported ISCC faces several paractical challenges as well as that mentioned in Section \ref{Sect:ChallengesISCC}. First, the long distance and environmental issues, such as bad weather conditions, atmospheric disturbances, electromagnetic interference, and physical obstacles may degrade the communication and remote sensing performance. Second, satellites have high mobility and the path loss between UAVs and satellites is large, leading to weak and varying communication channels. Next, each satellite has a wide coverage, resulting in large distance variations between the satellite and different connected UAVs. Consequently, communication time synchronization becomes a severe issue. As well, the battery life of UAVs and ground devices may limit their capabilities of sensing, communication, and computation. Moreover, the SAGIN has a large coverage to support a massive number of devices. Coordinating the sensing, communication, and computation processes for them is critically challenging. To this end, it calls for future research efforts to implement SAGIN-supported ISCC.


\subsection{Discussion of Unresolved Issues}

Although many efforts have been made toward ISCC as elaborated in Sections \ref{Sect:SignalDesign} and \ref{Sect:RRM}, there are many critical issues remaining unresolved to implement ISCC in the future. Some of them are discussed below.

\begin{itemize}
\item \emph{Lack of Theoretical Analysis Tools}: Although there are individual theoretical analysis for sensing, communication, and computation, and their partial integration e.g., ISAC and AirComp, there is a lack of theory that can characterize the integration performance of the three modules, especially from the information-theoretic perspective. This further results in a lack of guidance to design common ISCC strategies that can be generally applied in different tasks, scenarios, and networks. 

\item \emph{Hardware Design}: New hardware designs are required to enable resource sharing and coordination among sensing, communication, and computation modules. As an example, in an ISCC-supported online federated learning task, a pipeline design, where on-device computation can be conducted to execute stochastic gradient descent right after obtaining each sensory data sample, can enhance the time efficiency but calls for dedicated hardware that breaks up the barriers between sensing and computation.  

\item \emph{Protocols and Standardization}: Although many schemes and algorithms are proposed for ISCC, the designs of ISCC protocols and standardization remain uncharted. Particularly, an important issue is that the ISCC protocol should be the compatible with existing wireless communication systems. 

\item \emph{Security}: In some critical tasks, it is enssential to avoid information leakage during sensing, communication, and computation processes. However, the ISCC design enhances the correlation between the three processes. For example, it is possible to infer the sensory data information based on the communication and computation loads and the computational results. It is also possible to infer the computational task information according to the sensory view and modalities. As a result, more advanced security methods are needed in ISCC systems. 

\item \emph{Privacy}: In ISCC systems, the completion of a task generally requires sharing sensory of intermediate data between different devices, edge servers, or even clouds. Data privacy preservation become an important issue. On one hand, we can utilize the distortion in the ISCC systems, e.g., sensing/communication noise and quantization error, to preserve data privacy. On the other hand, it is challenging to preserve data privacy while guaranteeing good task performance. 

\item \emph{Robustness}: ISCC systems should be robust to unexpected device failure and the case where some functions (e.g., sensing or communication) of one or several devices get lost. Novel techniques are required to guarantee the robustness of ISCC systems. Besides, similar to the privacy issue, robustness enhancement may need the sacrifice of task performance. Balancing the tradeoff among the robustness, privacy, task performance, and other issues is important in the practical implementation of ISCC.

\item \emph{Backward Compatibility with Existing Technologies}: ISCC-enabled wireless systems should be compatible with existing wireless technologies. They should support old protocols and devices equipped with old hardware.

\end{itemize}
Overall, the efficient implementation of ISCC calls for many future research efforts to address the unresolved issues listed above.

\section{Conclusion}


ISCC, with the ability to improve network resource utilization and achieve customized goals of complicated tasks, turns to be a key technology for realizing connected intelligence in future networks. In this paper, we first provide a comprehensive survey in terms of the historical partial integration technologies, i.e., ICC, ISC, and ISAC. A holistic discussion on the shortages of these technologies is conducted to motivate ISCC, including low resource utilization, mismatch between goals of the overall task and each module (i.e., sensing, communication, or computation), and ignoring the tight coupling of different modules. Then, ISCC's benefits, functions in 6G for supporting the KPIs and usage scenarios, and challenges of ISCC are analyzed. Particularly, the benefits include sensing-communication-computation symbiosis, the ability to achieve higher network resource utilization and task performance. The challenges include multi-objective optimization, inference management, high signal design complexity, complicated resource management, lack of unified design criteria, and generalization. To tackle these challenges, the techniques of signal design and network resource management for ISCC are surveyed. Finally, the implementation of ISCC in several future advanced networks is investigated and the unresolved issues which call for future research efforts are discussed. Hopefully, this article will promote the advancement of ISCC in 6G, encouraging future research efforts in this interesting field.

\bibliography{reference}
\bibliographystyle{IEEEtran}

\end{document}